\documentclass[fleqn,usenatbib]{mnras}

\pdfoutput=1

\usepackage{newtxtext,newtxmath}
\usepackage[T1]{fontenc}

\DeclareRobustCommand{\VAN}[3]{#2}
\let\VANthebibliography\thebibliography
\def\thebibliography{\DeclareRobustCommand{\VAN}[3]{##3}\VANthebibliography}

\usepackage{subcaption}
\usepackage[export]{adjustbox}
\usepackage{orcidlink}
\usepackage{graphicx} 
\usepackage{amsmath}
\usepackage{multirow}
\usepackage{bigints}
\usepackage{soul}

\newcommand{\cov}{f_{\mathrm{cov}} (<R_{\mathrm{vir}})}

\newcommand{\covd}{f_{\mathrm{cov}} (r)}

\newcommand{\vir}{R_{\mathrm{vir}}}

\newcommand{\fb}{FIREbox}

\title[\ion{H}{i} covering fraction of LLSs in FIRE]
{The \ion{H}{i} covering fraction of Lyman Limit Systems in FIRE haloes}

\author[L. Tortora et al.]{
Lucas Tortora,$^{1, 2}$\orcidlink{0009-0005-8040-8325}\thanks{E-mail: lucas.tortora@uzh.ch}
Robert Feldmann,$^{1}$\orcidlink{0000-0002-1109-1919}
Mauro Bernardini$^{1}$\orcidlink{0000-0002-2930-9509}
and Claude-André Faucher-Giguère$^{3}$\orcidlink{0000-0002-4900-6628}
\\
$^{1}$Institute for Computational Science, University of Zurich, Winterthurerstrasse 190, Zurich CH-8057, Switzerland\\
$^{2}$Department of Physics, ETH Zurich, Otto-Stern-Weg 1, Zurich CH-8093, Switzerland\\
$^{3}$CIERA and Department of Physics and Astronomy, Northwestern University, 1800 Sherman Avenue, Evanston, IL 60201, USA
}

\date{Accepted 2024 July 15. Received 2024 June 17; in original form 2023 December 01}

\pubyear{2023}

\begin{document}
\label{firstpage}
\pagerange{\pageref{firstpage}--\pageref{lastpage}}
\maketitle

\begin{abstract}
Atomic hydrogen (\ion{H}{i}) serves a crucial role in connecting galactic-scale properties such as star formation with the large-scale structure of the Universe. While recent numerical simulations have successfully matched the observed covering fraction of \ion{H}{i} near Lyman Break Galaxies (LBGs) and in the foreground of luminous quasars at redshifts $z \lesssim 3$, the low-mass end remains as-of-yet unexplored in observational and computational surveys.
We employ a cosmological, hydrodynamical simulation (\fb{}) supplemented with zoom-in simulations (MassiveFIRE) from the \textit{Feedback In Realistic Environments} (FIRE) project to investigate the \ion{H}{i} covering fraction of Lyman Limit Systems ($N_{\ion{H}{i}} \gtrsim 10^{17.2}$ cm$^{-2}$) across a wide range of redshifts ($z=0-6$) and halo masses ($10^8-10^{13} \, M_{\sun}$ at $z=0$, $10^8-10^{11}\, M_{\sun}$ at $z=6$) in the absence of feedback from active galactic nuclei. 
We find that the covering fraction inside haloes exhibits a strong increase with redshift, with only a weak dependence on halo mass for higher-mass haloes. For massive haloes ($M_{\mathrm{vir}} \sim 10^{11}-10^{12} M_{\sun}$), the radial profiles showcase scale-invariance and remain independent of mass. The radial dependence is well-captured by a fitting function. The covering fractions in our simulations are in good agreement with measurements of the covering fraction in LBGs. Our comprehensive analysis unveils a complex dependence with redshift and halo mass for haloes with $M_{\mathrm{vir}} \lesssim 10^{10} M_{\sun}$ that future observations aim to constrain, providing key insights into the physics of structure formation and gas assembly. 
\end{abstract}




\section{Introduction}

In the concordance $\Lambda$CDM cosmology, the large-scale structure of the Universe results from the collapse of a collisionless, cold dark matter and subsequent clustering of baryons. Cosmological simulations show that initial dark matter overdensities hierarchically assemble into virialized structures called dark matter haloes which are interconnected into a filamentary structure known as the cosmic web \citep[see][]{1980lssu.book.....P, 1983FCPh....9....1E}{}. Baryons later fall into the potential wells mapped out by these haloes and become gravitationally bound to them. The dark matter haloes hence form the building blocks for large-scale structure formation in which galaxies and clusters of galaxies are born \citep[e.g., ][]{1978MNRAS.183..341W, 1991ApJ...379...52W, 2010MNRAS.404.1111G, 2018ARA&A..56..435W}{}

Galaxies continue interacting with gas from the intergalactic medium (IGM) during their lifetime, within an intricate combination of accretion and feedback processes. To sustain their growth, they need to obtain fresh gas from the IGM  \citep[e.g.][]{2005MNRAS.363....2K, 2010ApJ...717..323B}. The nature of this supply is strongly dependent on redshift and halo mass. For instance, the gas is shock-heated in massive haloes and takes a long time before settling in the galactic disk \citep[e.g.,][]{1977ApJ...215..483B, 1977MNRAS.179..541R, 2003MNRAS.345..349B}. In less massive haloes, much of the accretion occurs instead via the cold mode \citep[e.g.,][]{2009Natur.457..451D, 2011MNRAS.417.2982F, 2019ApJ...875...54H, 2020MNRAS.492.6042S}.
This gas can collapse into molecular clouds which then become stellar nurseries \citep[e.g.,][]{1965PThPh..34..754H, 2002ApJ...564...23B, 2019NatAs...3.1115D}, or fall into the center of galaxies and ignite active galactic nuclei \citep[AGN; e.g.][]{1993ARA&A..31..473A, 1995PASP..107..803U, 2017A&ARv..25....2P}.  
Feedback processes regulate star formation and gas accretion by launching powerful outflows into the surrounding regions of these galaxies, affecting their dynamics and morphology \citep[see][]{2017ARA&A..55..389T, 2018MNRAS.480..800H, 2018MNRAS.475.5688B, 10.1093/mnras/stz3131,2023ARA&A..61..131F}. The complex distribution and physics of gas around galaxies thus contains the fingerprint of galactic properties. Studying the absorption features of elements such as hydrogen and heavier metals in the spectra of bright background sources allows us to understand the principal physical processes governing galactic properties.

Advances in both simulations and observational campaigns have led to significant advances in comprehending stellar properties and molecular gas within galaxies \citep[e.g.][]{2015MNRAS.450.2749G, 2016ApJ...833...68A, 2020A&A...643A...1L, 2020ARA&A..58..157T, 2020CmPhy...3..226F}. 
Much is still unknown about atomic hydrogen \ion{H}{i}, specifically at higher redshifts, due to its lack of allowed transitions at cold temperatures resulting in a challenging detection process.
As the most abundant element in the Universe, mapping its distribution through cosmic time promises to offer crucial constraints on galactic evolution and cosmology \citep[see e.g.,][]{padmanabhan_2017, Dutta_2019}. In the low redshift Universe ($z \lesssim 1$), the (highly forbidden) 21cm line can be used to directly map the \ion{H}{i} distribution \citep[e.g.,][]{2012MNRAS.420.2924K, 2015MNRAS.450..926R}. Current and future observing campaigns with improved sensitivity, such as MeerKAT \citep{2009arXiv0910.2935B, 2016mks..confE...1J} and SKA \citep{2020PASA...37....2W}, will use the 21cm emission to also map the distribution of hydrogen at higher redshifts. At these redshifts, absorption lines in the spectra of bright background sources are generally used to investigate the presence of atomic hydrogen along specific sight-lines \citep[see e.g.,][]{2011ApJ...737L..37A, 2019MNRAS.489.4926G}. 
These studies showed that \ion{H}{i} can be found across a large range in column densities, including Lyman Limit Systems (LLSs; with column density $N_{\ion{H}{i}} > 10^{17.2}$ cm$^{-2}$), see e.g. \citet{2012A&A...547L...1N, 10.1093/mnras/stv1182, padmanabhan_2017}. 

Current simulations predict that a substantial fraction of accreted material that enters haloes is relatively cold and therefore could contain considerable amounts of neutral gas \citep[][]{2011MNRAS.418.1796F, 2011MNRAS.412L.118F, 2012MNRAS.421.2809V, 2013MNRAS.429.3353N, 2014ApJ...780...74F}. Furthermore, the number and column density of \ion{H}{i} absorbers is predicted to increase closer to galaxy centers, suggesting that absorbers with high \ion{H}{i} column density are better probes of gas in the proximity of galaxies \citep[e.g.,][]{2015MNRAS.452.2034R, 2019MNRAS.487.1529D, 2021MNRAS.507.2869S}. However, strong \ion{H}{i} absorbers such as LLSs are predicted to be often close to galaxies that may be too faint to be detected in actual surveys \citep[][]{2014MNRAS.438..529R}. The study of strong \ion{H}{i} absorbers around bright galaxies residing in massive haloes ($\geq 10^{12} M_{\sun}$) can help to overcome this problem. In fact, many modern observations and simulations make use of this galaxy-centered approach to measure covering fractions of neutral hydrogen clouds \citep[e.g.][]{2009ApJ...701.1219C, 2012ApJ...751...94R, 2013ApJ...777...59T, 2013ApJ...776..136P, 2014MNRAS.445..794T, 2015ApJ...808...38R, 2017ApJ...837..169P}. 

The observational constraints have motivated several groups to investigate the distribution of \ion{H}{i} surrounding galaxies via the use of simulations \citep[for instance,][]{2011MNRAS.418.1796F, 2014ApJ...780...74F,  2013ApJ...765...89S, 2015MNRAS.449..987F, 2015MNRAS.453..899M, 2017MNRAS.468.1893M, 2017MNRAS.464.2796G, 2019MNRAS.483.4040S, 2020MNRAS.498.2391N, 2021MNRAS.501.4396G, 2021MNRAS.507.2869S, weng2023physical}. 
Historically, the high observed covering fractions reported by \citet{2012ApJ...750...67R} around Lyman Break Galaxies (LBGs) and by \citet{2013ApJ...762L..19P, 2013ApJ...776..136P} in the vicinity of quasars (QSOs) have been a challenge to reproduce in cosmological simulations \citep[see e.g.][]{2011MNRAS.412L.118F, 2014ApJ...780...74F, 2015MNRAS.449..987F}. 
These simulations are based on zoom-ins that focus on one galaxy \citep{2011MNRAS.412L.118F, 2013ApJ...765...89S} or include only a few galaxies with a limited range of masses and redshifts \citep{2014ApJ...780...74F, 2015MNRAS.449..987F}. There are several caveats to these approaches. For instance one could expect, given the diversity of the observed objects, that a large sample of simulated galaxies is required to accurately compare with observations of the \ion{H}{i} distribution \citep{2015MNRAS.452.2034R}. Furthermore, other constraints such as the cosmic distribution of \ion{H}{i} should be satisfied.
Finally, constraints on current instrumentation limit observations to massive haloes of over $10^{12} M_{\sun}$, hence requiring a large volume or a high number of zoom-ins to be able to obtain a meaningful statistical distribution of these systems in simulations \citep[][]{2011ApJ...737L..37A, 2020MNRAS.494.1143B}.

Past works have demonstrated that accurately replicating the properties of the circumgalactic medium (CGM) and the resulting gas covering fractions of galaxies is complicated by the effects of resolution and the details and implementation of both star formation and feedback mechanisms \citep[e.g.][]{2015MNRAS.449..987F, 2015MNRAS.448..895S, 2015MNRAS.452.2034R, 2019MNRAS.482L..85V, 2020MNRAS.499.2760S}. 
Resolution is particularly critical, as the quantity of cold gas in the CGM is not converged \citep[see e.g.][]{2023ARA&A..61..131F, ramesh2023zooming}. Although the precise details vary, simulations with a generally stronger stellar feedback implementation have produced consistently higher values of covering fraction \citep{2015MNRAS.449..987F, 2016MNRAS.461L..32F, 2015MNRAS.452.2034R}, and hence reconciled the results found by \citet{2012ApJ...750...67R} around LBGs, as opposed to works with weaker feedback \citep{2011MNRAS.412L.118F, 2014ApJ...780...74F}. Observational data from the COS-Halos survey \citep{2013ApJ...777...59T, 2017ApJ...837..169P} and simulations \citep[][]{2015MNRAS.449..987F, 2015MNRAS.452.2034R} show little correlation between the instantaneous star-forming activity or specific star-formation rate (sSFR) and column density of \ion{H}{i}. On the other hand, the precise role of AGN feedback in shaping the CGM cool gas distribution remains open to discussion, with the existing literature citing either insignificant or important effects when including them \citep{2015MNRAS.452.2034R, weng2023physical, 2024MNRAS.529..537K}. While AGNs are ubiquitous in reality and their various feedback mechanisms are necessary to construct a full picture of galaxy formation, how to best model them in cosmological simulations remains uncertain, and the current state-of-the-art models necessarily introduce modelling degeneracies which significantly hinder predictive power.

The most recent studies on the covering fraction of \ion{H}{i} have produced a robust agreement with results from both the LBGs and QSOs observations, mainly by remedying the hindering factors mentioned above and studying a great number of simulated objects in large cosmological volumes and concluding on the average distribution of \ion{H}{i}. \citet{2015MNRAS.452.2034R} surveyed the covering fraction of atomic hydrogen over a large range of masses ($M_{200} \in [10^{11} - 10^{13.7}] M_{\odot}$) in the full-scale cosmological simulations EAGLE \citep{2015MNRAS.450.1937C, 2015MNRAS.446..521S}, and found an agreement with the radial distribution of \ion{H}{i} around QSOs reported by \citep{2013ApJ...776..136P}. Additionally, \citet{2016MNRAS.461L..32F} complemented their previous works with multiple higher resolution zoom-ins and found their results for the most massive haloes in their simulations ($M_{\mathrm{vir}} \gtrsim 10^{12.5}M_{\sun}$) to also be consistent with the covering fractions observed around quasars.\\

In this work, we offer a description of covering fractions of Lyman Limit Systems using the high-resolution cosmological \fb{} simulation \citep{2023MNRAS.522.3831F} and a series of zoom-ins from the MassiveFIRE suite \citep{2016MNRAS.458L..14F, 2017MNRAS.470.1050F} rerun with FIRE-2 physics \citep{2017MNRAS.470.4698A}. This allows us to resolve haloes with very-low mass (from $M_{\mathrm{vir}} = 10^{7.75} M_{\sun}$) and investigate the profiles of covering fraction from a much lower mass range than that of other simulations. The analysis is done for redshifts $0 \leq z \leq 6$, offering a more consistent investigation of the redshift dependence of covering fractions. The properties of these simulations mean we have the high resolution necessary to resolve the small-scale gas structure around haloes, while also ensuring that we have a large enough sample of haloes to obtain systematic information about them. 

The paper is structured as follows. In section ~\ref{method}, we introduce the cosmological simulations and the methodology for our analysis. In section ~\ref{results} we present the relevant results and discuss our findings. We discuss our results in the context of other works and observations in section ~\ref{discussion}. We finally conclude in section ~\ref{conclusions}.

\section{METHODOLOGY} \label{method}

\subsection{Simulations}

\fb{} is a high-resolution hydrodynamic cosmological volume simulation with a box size of 22.1 cMpc \citep{2023MNRAS.522.3831F}. \fb{} is part of the \textit{Feedback In Realistic Environments} (FIRE)\footnote{The official FIRE project website can be found here: \url{https://fire.northwestern.edu}} project \citep{2014, 2018MNRAS.480..800H, 2023MNRAS.519.3154H}. In the next paragraphs, we briefly discuss the details of the simulation.

The simulation volume contains 1024$^3$ dark matter particles and 1024$^3$ gas particles at the initial redshift ($z=120$). Dark matter and baryon masses are $m_{\mathrm{DM}} = 3.35 \times 10^5 M_{\sun}$ and $m_{\mathrm{b}} = 6.26 \times 10^4 M_{\sun}$ respectively. Dark matter (star) particles have a fixed softening length of 80 pc (12 pc). Gas softening is adaptive with a minimum softening length of 1.5 pc.
Initial conditions were generated with the \textsf{MUSIC} (MUlti-Scale Initial Conditions) code \citep{2011} and with 2015 Planck cosmological parameters \citep{2016}: $H_0 = 67.74$ km/s/Mpc (or $h = 0.6774$), $\Omega_{\mathrm{m}} = 0.3089$, $\Omega_{\Lambda} = 0.6911$ , $\Omega_{\mathrm{b}} = 0.0486$ , $\sigma_8 = 0.8159$ and $n_s = 0.9667$.  

\fb{} is run with the gravity-hydrodynamics solver \textsf{GIZMO} \citep{10.1093/mnras/stv195}\footnote{A public version of the code is available at: \url{http://www.tapir. caltech.edu/~phopkins/Site/GIZMO.html}} using the FIRE-2 physics model \citep{2018MNRAS.480..800H}. 
The simulation incorporates multiple gas-cooling processes (such as: free-free, photo-ionization, recombination, Compton, photoelectric, metal-line, molecular, fine-structure, dust collisional, and cosmic ray physics) following an implicit algorithm described in \cite{2018MNRAS.480..800H}. FIREbox includes radiative feedback in the form of photo-ionization and photoelectric heating. Radiative transfer effects are accounted for in our simulations via the Locally Extincted Background Radiation in Optically
thin Networks approximation \citep[LEBRON; ][]{2012MNRAS.421.3488H, 2014, 2018MNRAS.480..800H, 2019MNRAS.483.4187H}. To compute the \ion{H}{i} fraction in each resolution element, the model assumes equilibrium between recombination, collisional ionization and photoionization by local stellar sources. The process of self-shielding is taken into account using a local Sobolev/Jeans-length approximation which is calibrated from radiative transfer experiments \citep{2010ApJ...725..633F, 2013MNRAS.431.2261R}. Approximating the attenuation of incident flux in this manner was shown to reproduce neutral fractions calculated with full radiative transfer codes in post-processing \citep[see][for further details]{2010ApJ...725..633F, 2015MNRAS.449..987F, 2013MNRAS.431.2261R, 2021MNRAS.507.2869S}. Relevant metal ionization states are tabulated from the \textsf{CLOUDY} simulations \citep{1998PASP..110..761F}. Feedback from AGNs is not included. We refer the interested reader to \citet{2018MNRAS.480..800H} for the full description of the FIRE-2 simulations.

To offer a more complete comparison of our results with observing campaigns, we supplement our analysis of the higher mass end of haloes with four zoom-in simulations (A1, A2, A4, A8) from the MassiveFIRE \citep{2016MNRAS.458L..14F, 2017MNRAS.470.1050F} suite, re-simulated with FIRE-2 physics \citep{2017MNRAS.470.4698A}. We briefly outline the selection criteria of the zoom-ins. Isolated haloes with $M_{\mathrm{vir}} \, (z=2) \approx 10^{12.5} M_{\sun}$ were selected from low-resolution, dark matter-only runs. The environmental density of each halo was then measured by computing the total mass in a sphere of radius 1.8 Mpc. The final haloes of the zoom-in suite were selected from the 5th, 25th, 50th, 75th and 95th percentile of the distribution of environmental density. This choice results in the haloes having varied accretion histories \citep[see][Section 2, for details]{2017MNRAS.470.1050F}. The masses of particles are $m_{\mathrm{DM}} = 1.76 \times 10^5 M_{\sun}$ and $m_{\mathrm{b}} = 3.29 \times 10^4 M_{\sun}$, and dark matter (gas) particles have softening lengths of 143 pc (9 pc).

\subsection{Covering fractions}
Our goal is to compute the \ion{H}{i} covering fraction of the simulated haloes. We define this quantity and outline the procedure to obtain column density maps from simulation snapshots in the next paragraphs. \\

The covering fraction is used to quantify the distribution of atomic hydrogen around haloes. In our study, we mainly consider the covering fraction of strong \ion{H}{i} absorbers, specifically LLSs (with $N_{\ion{H}{i}} > 10^{17.2}$cm$^{-2}$) without further separating them from Damped Lyman-$\alpha$ systems (with $N_{\ion{H}{i}} > 10^{20.3}$cm$^{-2}$). 
We use two separate measures to quantify the \ion{H}{i} distribution around haloes: the cumulative and differential covering fractions, respectively denoted by $f_{\mathrm{cov}} (< R)$ and $f_{\mathrm{cov}} (r)$. They are related according to the following formula:
\begin{equation} \label{fcov-rel}
    f_{\mathrm{cov}} (< R) = \frac{\int_0^R f_{\mathrm{cov}} (r) \cdot 2\pi r \cdot dr}{\pi \cdot R^2} \equiv \frac{A_{\mathrm{abs}} \, (< R)}{\pi \cdot R^2}
\end{equation}
where $A_{\mathrm{abs}} \, (< R_{\mathrm{vir}})$ is the area covered by LLSs within the field of view defined by the virial radius of a given halo. The cumulative covering fraction of LLSs essentially measures the probability of finding such systems in line-of-sights within some radius $R$. \\
The differential covering fraction is used to quantify the spatial distribution of LLSs around haloes and complements the previous measure. Effectively, it is computed according to:
\begin{equation}
    \covd = \frac{A_{\mathrm{abs}} \, (r_i < r < r_{i+1})}{\pi \cdot (r_{i+1}^2 - r_i^2)}
\end{equation}
with $A_{\mathrm{abs}} \, (r_i < R < r_{i+1})$ being the area covered by LLSs within an annulus of inner and outer impact parameters $r_i$ and $r_{i+1}$ respectively. The $r_i$'s are consistently chosen such that $r_0 = 0$, and we decided to use more annuli closer to the center of the halo. This choice is motivated by previous works and ensures we can capture details of the shape of the differential profile.

\subsection{Data generation}
We use the snapshots corresponding to $z = 0, 1, 2, 2.5, 3, 4, 5, 6$ from the \fb{} simulation for our study. In order to obtain covering fractions, we compute column density maps by depositing particle densities from the simulated cube onto a uniform grid using the \texttt{smooth} and \texttt{tipgrid} algorithms.
For each particle, \texttt{smooth} calculates a smoothing length specified as half the distance to its $n^{\mathrm{th}}$ neighbour particle. We set $n = 10$ for this work. We note that lower values of $n$ translate into higher particle noise but allow us to better resolve small structures. Following this, we subdivide our simulated box into 20 equally spaced slabs (of thickness $\sim$1.1 cMpc), along each of the 3 spatial directions. The \texttt{tipgrid} algorithm then interpolates particles within the same slab onto a two-dimensional grid, by distributing the mass using a spherically-symmetric kernel according to the aforementioned smoothing lengths. We chose a grid resolution of 32768$^2$ pixels, such that individual pixels resolve roughly 0.68 ckpc. This choice ensures we can observe the finer details in the structures formed in the simulation.

We identify simulated haloes and recover their main properties using the \textsf{AMIGA} Halo Finder \citep[AHF;][]{2004MNRAS.351..399G, 2009ApJS..182..608K}. We include all haloes with at least 168 dark matter particles in our analysis. This means that all AHF haloes with roughly $M_{\mathrm{vir}} \gtrsim 10^{7.75} M_{\sun}$ are included, which enables us to study the behaviour for much smaller structures and masses than done in previous works. The virial mass ($M_{\mathrm{vir}}$) and virial radius ($\vir$) of dark matter haloes are computed following the virial overdensity definition outlined in \citet{1998ApJ...495...80B}. Using the AHF particle files, we identify haloes by their center and virial radius and distinguish between ‘main haloes' and ‘sub-haloes'. In this work, the term `sub-haloes' refers to dark matter haloes that are nested within another dark matter halo. All other haloes identified by AHF are `main haloes'. For sub-haloes, the virial radius $\vir$ reported by AHF and used throughout this work refers to the smaller of the virial and tidal radius \citep[see, e.g.,][]{vdbosch+2018}. In particular, the tidal radius is significantly smaller than the virial radius for sub-haloes close to the center of the host main halo.

There are roughly 96000 haloes at redshift $z=6$, growing to a peak of 160000 haloes at redshift $z=2$ and finally 130000 at $z=0$. The main-to-sub proportion evolves from 95\%-5\% at $z=6$ to 80\%-20\% at $z=0$. This selection offers a statistically sound sample of covering fractions for all halo masses below some redshift-dependent higher-mass of $M_{\mathrm{vir}} \sim 10^{11} M_{\sun}$ at redshift $z=6$, up to $M_{\mathrm{vir}} \sim 10^{13} M_{\sun}$ at redshift $z=0$.

For our study, we always evaluate the covering fraction in Eq. \eqref{fcov-rel} for $R = \vir$. It is computed for each halo according to the following straightforward procedure. We place a circular mask around haloes so that only pixels within the virial radius are considered. We then place another circular mask so that only those pixels with column density above the threshold are counted. The ratio of the two gives the covering fraction. The differential covering fraction is obtained in the same way, with annuli being used instead of disks. We repeat this procedure for all redshifts and for all 3 orientations of the simulated cube, and compute statistics using each orientation as an independent data point for our final results.

\section{RESULTS} \label{results}

\subsection{Cumulative covering fraction of LLSs}

\begin{figure*}
	\centering
	\includegraphics[width=\linewidth]{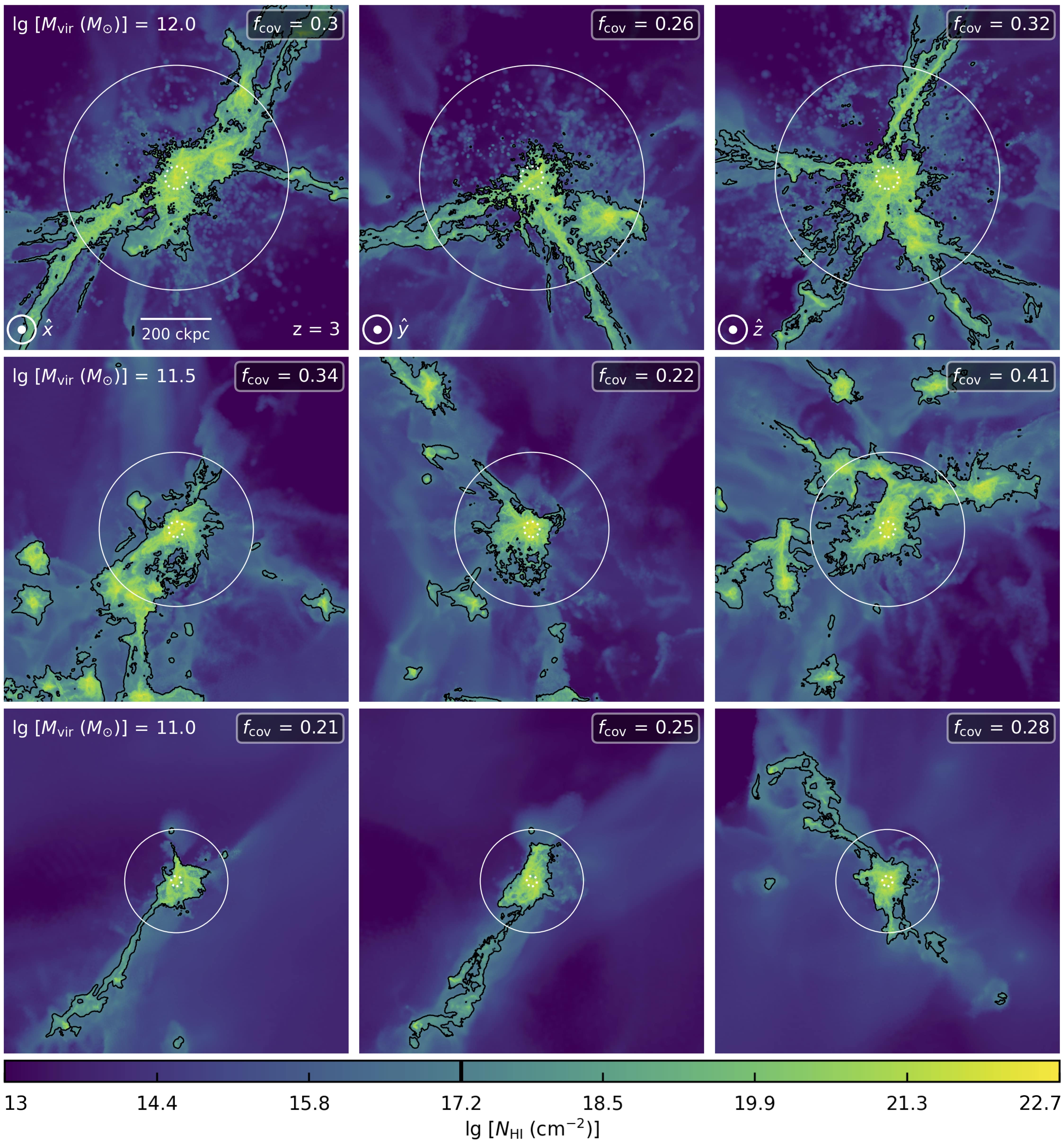}
    \caption{\ion{H}{i} column density distribution in surroundings of randomly selected main haloes, at redshift $z = 3$. Top, middle and bottom rows show haloes with $M_{\mathrm{vir}} \approx 10^{12}, 10^{11.5}$ and $10^{11} \, M_{\sun}$ respectively. The columns in each row show the same halo from three different orthogonal projections. The panels show a region of size 1 $\times$ 1 cMpc$^2$, with the same depth of projection. The full and dotted circles in white are centred on the haloes and display their virial radius $R_{\mathrm{vir}}$ and 10\% of their virial radius, respectively. Black contours highlight LLS sightlines (with $N_{\ion{H}{i}} > 10^{17.2} \mathrm{cm}^{-2}$), and the covering fraction of LLSs within $R_{\mathrm{vir}}$ is indicated in the top-right of each panel. The covering fraction does not evolve strongly with mass for $M_{\mathrm{vir}} \geq 10^{11} M_{\sun}$ haloes, but can significantly differ from one orthogonal projection to another of the same halo.} 
    \label{pic_masses_same}
\end{figure*}

\begin{figure*}
	\centering
	\includegraphics[width=\linewidth]{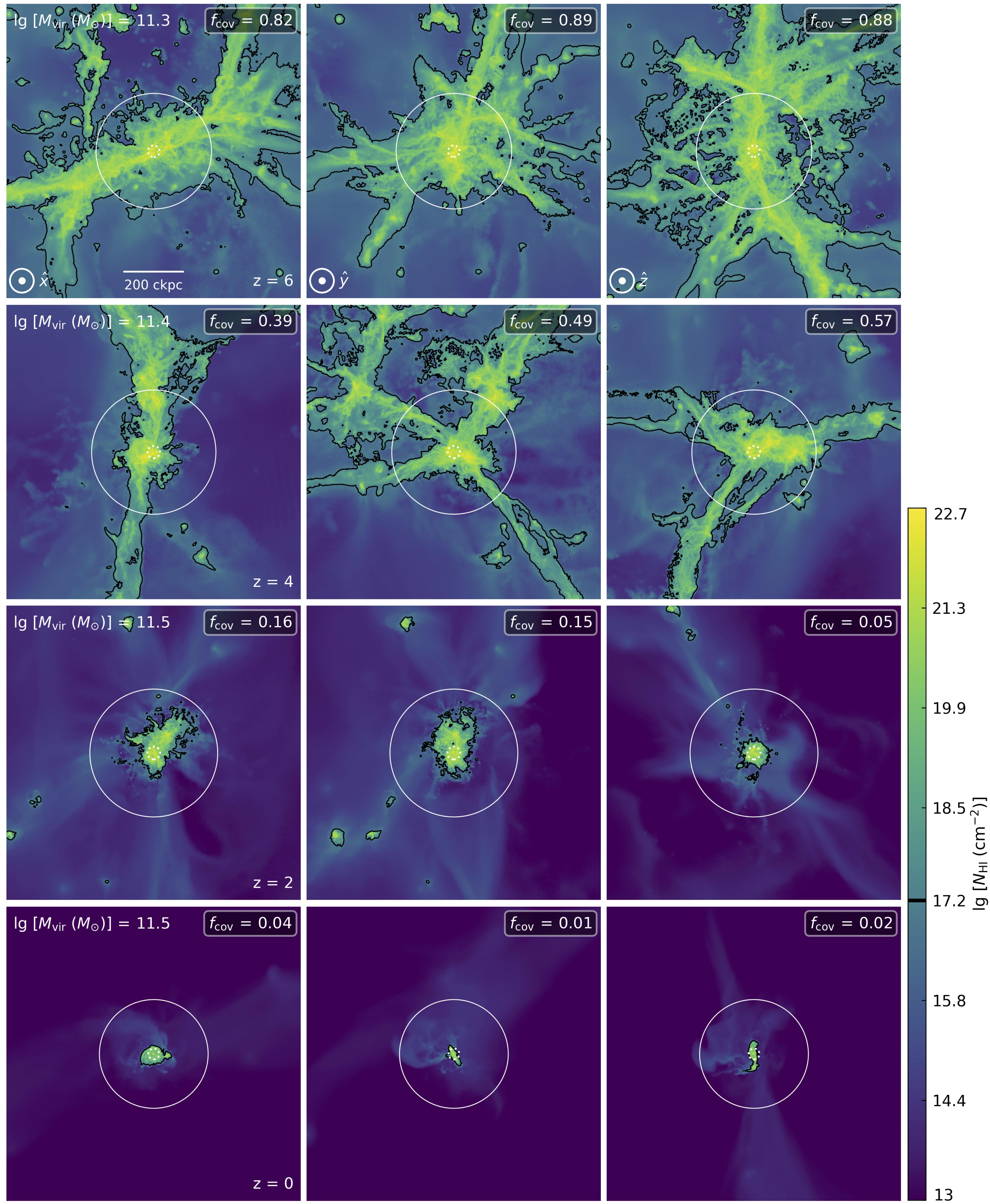}
    \caption{\ion{H}{i} column density distribution in surroundings of randomly selected main haloes. The rows show haloes with $M_{\mathrm{vir}} \approx 10^{11.5} M_{\sun}$ at redshifts $z=6, 4, 2$ and $0$ (respectively, from top to bottom). The columns in each row show the same halo from three different orthogonal projections. The panels show a region of size 1 $\times$ 1 cMpc$^2$, with the same depth of projection. The full and dotted circles in white are centred on the haloes and display their virial radius $R_{\mathrm{vir}}$ and 10\% of their virial radius, respectively. Black contours highlight LLS sightlines (with $N_{\ion{H}{i}} > 10^{17.2} \mathrm{cm}^{-2}$), and the covering fraction of LLSs within $R_{\mathrm{vir}}$ is indicated in the top-right of each panel. The filamentary structure of \ion{H}{i} in and around FIREbox haloes evolves very strongly with redshift, from almost complete coverage of the virial cross-section at redshift $z=6$ to being highly centrally contracted within haloes at redshift $z=0$. The covering fraction thus increases significantly with increasing redshift, and we find that this is the predominant parameter affecting values of $\cov$.}
    \label{pic_redshift_same}
\end{figure*}

In the following paragraphs, we begin investigating the covering fraction of Lyman Limit Systems in FIREbox haloes by directly taking a look at some snapshots from the simulations and performing a visual inspection of the pictures. Figures ~\ref{pic_masses_same} and ~\ref{pic_redshift_same} show examples of the distribution of \ion{H}{i} around randomly-chosen main haloes, with either changing mass or redshift. 

Fig.~\ref{pic_masses_same} shows haloes of different mass from different orthogonal projections at redshift $z=3$. We see that the gas distribution is highly inhomogeneous, with a filamentary structure that can extend beyond the virial radius of the haloes. The shape and extent of this structure also change significantly when viewing the same halo from different angles. 

We note that a significant fraction of the projected area within one virial radius of the center of the haloes is covered by LLSs. The covering fraction of the haloes does not seem to be strongly correlated with virial mass: the gas around $M_{\mathrm{vir}} = 10^{12} M_{\sun}$ haloes covers a much greater area than that of $M_{\mathrm{vir}} = 10^{11} M_{\sun}$ haloes and the virial radius of $10^{12} M_{\sun}$ haloes is also much bigger than that of $10^{11} M_{\sun}$ haloes; the two effects are comparable in magnitude so that $\cov$ is about the same for these haloes. The inhomogeneity of the gas distribution noted above results in different values of covering fraction for different projections of the same halo. Specifically, the individual value of \ion{H}{i} covering fraction of a halo can vary up to a factor of $\sim 2$ from one angle of viewing to another. 

The pictures can also be used to visually highlight any redshift dependence of the distribution of neutral hydrogen around haloes. Fig. ~\ref{pic_redshift_same} shows haloes with mass $M_{\mathrm{vir}} \approx 10^{11.5} M_{\sun}$ at 4 different redshifts used in the study for three different orthogonal projections each. We find that the filamentary structure described in the previous figure evolves very strongly with redshift. At $z=6$ we observe large filaments and clumps of atomic gas that extend far beyond the virial radius of the studied haloes, interlinking huge regions of space. Such structure diminishes in extent and compactness with redshift, until they have almost disappeared by $z=0$. This is found to be consistent with the reports that at $z \lesssim 2$, essentially all the \ion{H}{i} is found inside dark matter haloes \citep[e.g.,][]{2018ApJ...866..135V, 2023MNRAS.522.3831F}. Consequentially, the covering fraction increases very strongly with increasing redshift: only around 3\% of a halo's virial radius is covered in LLSs at $z = 0$, whereas almost 90\% is covered at $z=6$. We find this redshift relation to be the predominant parameter in determining the covering fraction of \ion{H}{i} gas for a halo.

\begin{figure*}
	\includegraphics[width=\linewidth]{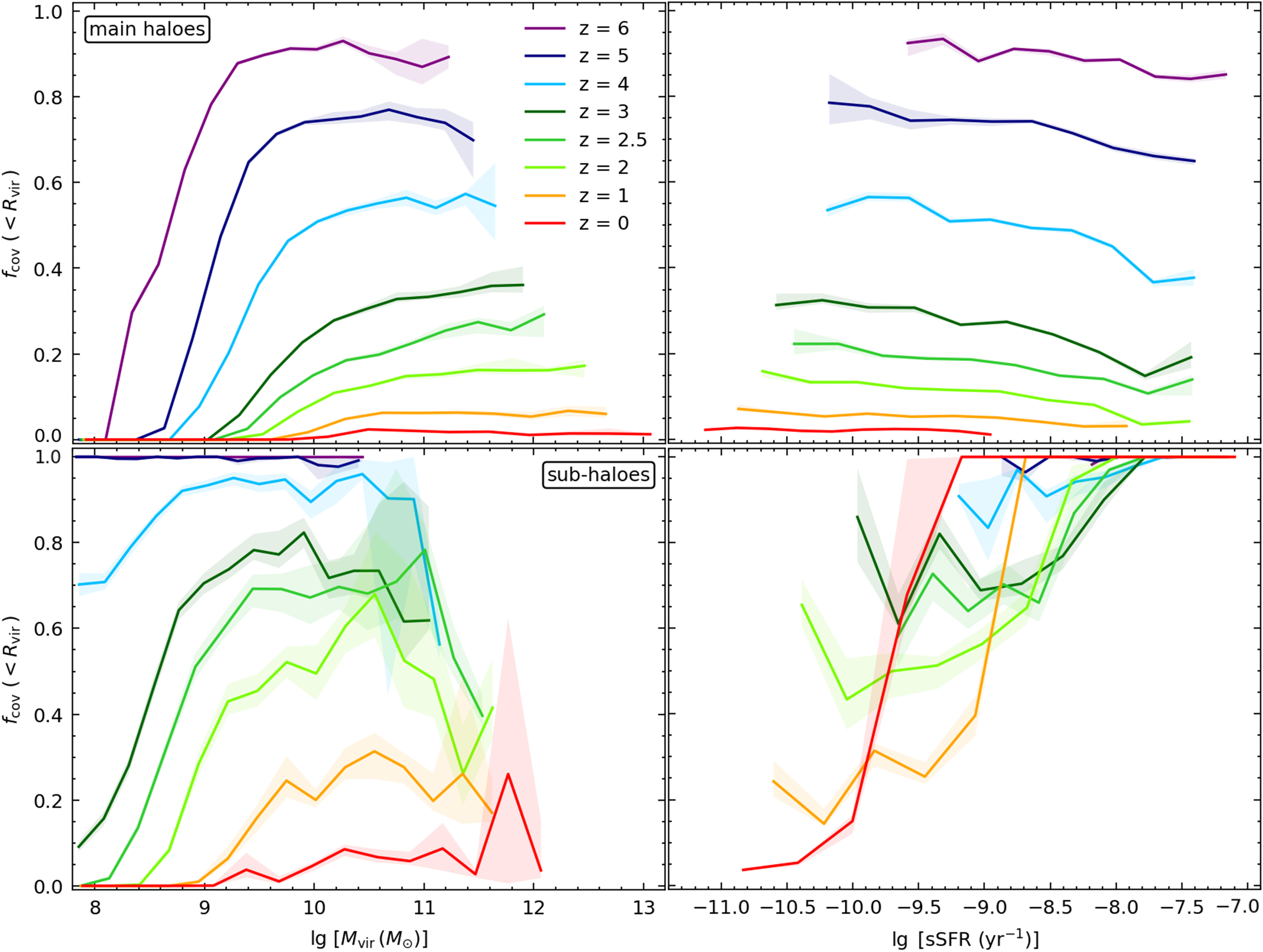}
    \caption{Cumulative covering fraction of LLSs within impact parameter $\vir$ as a function of halo mass $M_{\mathrm{vir}}$ (left column) and specific star formation rate (sSFR) $\dot{M}_{\star} / M_{\star}$ (right column). We distinguish between main- (top row) and sub- (bottom row) haloes. Different colours are used to distinguish between redshift from $z=0$ to $z=6$ as per the legend. Solid lines indicate the median covering fraction and shaded areas show the 5th-95th percentile error on the median obtained from bootstrapping. All haloes with $M_{\mathrm{vir}} \geq 10^{7.75} M_{\sun}$ identified in the simulation were used to create these plots. The behaviour of main- and sub-haloes is very different and details are discussed in the main text. The covering fraction of LLSs increases strongly with redshift and (for low-mass haloes) with halo mass. The covering fraction is (weakly) anticorrelated with sSFR in main haloes.}
    \label{cflls}
\end{figure*}

We proceed to a systematic statistical study of covering fractions over a wider range of masses and redshifts. Figure ~\ref{cflls} shows the measured $\cov$ for all sampled haloes at $z=0-6$. The covering fraction is studied as a function of halo virial mass $M_{\mathrm{vir}}$ and specific star-formation rate $\dot{M}_{\star} \, / \, M_{\star}$, for all resolved haloes with $M_{\mathrm{vir}} > 10^{7.75} M_{\sun}$. We distinguish between main- and sub-haloes. The haloes are grouped in equidistant mass bins, and for each one we show the median covering fraction and highlight the 5th-95th percentile error on the median found via bootstrapping. We find that the results significantly change between main- and sub-haloes.

For the main haloes (top row of Fig. ~\ref{cflls}), we conclude that the median covering fraction of LLSs strongly increases with increasing redshift, at all halo masses. We note that there is a particularly noticeable rise from $z=2$ to $z=3$ and all redshifts thereafter, as compared to the evolution between redshifts $z=0, 1$ and 2.\footnote{Redshift $z=2.5$ is introduced specifically to compare with observational surveys (see section ~\ref{discussion}).} Our results are consistent with the idea that haloes contain higher gas fractions at high redshift, due to increased accretion of gas and a higher mean density of the Universe \citep[see e.g.][]{2015MNRAS.452.2034R}. 
This finding is of particular importance for comparisons with observations. Indeed, since observed samples contain galaxies with a wide range of redshifts, the observed probability of finding an LLS sightline within a given impact parameter is not directly equal to the covering fraction of LLSs at the mean redshift of the sample, because higher-redshift galaxies contribute more to the covering fraction than lower-redshift ones \citep{2015MNRAS.452.2034R}. The predicted scatter of covering fractions at all redshifts further accentuates this issue and magnifies errors, underscoring that even minor inaccuracies in redshift estimation can lead to significant misjudgements of covering fractions from observations. Finally, we should mention that the virial radius of haloes cannot be directly observed and predominantly serves as a parameter in theoretical and numerical studies, meaning that $\cov$ as shown in Fig. ~\ref{cflls} is inherently challenging to compare with observations.

The complex halo-mass dependence of the covering fraction of main-haloes can also be studied with Fig. ~\ref{cflls}. At lower masses, the covering fraction is close to zero until some threshold mass between $10^{8} M_{\sun}$ and $10^{10} M_{\sun}$ depending on redshift. The covering fraction then rapidly increases with halo mass, at all redshifts. At higher masses, the covering fraction does not evolve strongly with halo mass and eventually plateaus at some value, which is higher for higher redshifts. We conclude that the covering fraction of massive ($M_{\mathrm{vir}} \approx 10^{11} - 10^{13} M_{\sun}$, depending on redshift) haloes is nearly independent of halo mass, at any given redshift. We conduct a deeper analysis of this mass-dependence when studying the \ion{H}{i} differential covering fraction of haloes in the following subsection.

As the top-right panel of Fig. ~\ref{cflls} highlights, the covering fraction is weakly anticorrelated with sSFR in main haloes. Together with results from \citet{2015MNRAS.449..987F}, this suggests that the covering fraction is not affected by the details of star formation, namely the instantaneous star formation rate, in haloes. On longer timescales, as highlighted previously, stellar feedback is very important for shaping the CGM and its properties such as the covering fraction of atomic gas \citep[see e.g.][]{2016MNRAS.461L..32F}. This argument more easily enables comparisons of predicted covering fractions because they do not need to be compared with galaxies in the exact same stages of star formation.
This immediate result also strengthens our conclusion that covering fraction strongly increases with redshift and that this is the main parameter with which it varies.

For the sub-haloes (bottom row of Fig. ~\ref{cflls}), it is also found that $\cov$ increases with redshift. We note that the covering fraction is generally higher for sub-haloes than main haloes, albeit with significant scatter, at the same virial mass. This is likely due to a geometric effect, whereby the sub-haloes can be entirely covered by gas found within the virial radius of more massive main haloes. As such, some sub-haloes can be absorbed ‘into' or pushed ‘out' of the field of view of the main halo when viewed from different orientations. This usually results in a notable increase of $\cov$, but also produces significant scatter as is seen by the large shaded areas around the median in Fig. ~\ref{cflls}. The scatter can be explained by a combination of both the randomness of projection effects discussed above and the smaller sample size ($\sim 4-20$ times fewer sub- than main-haloes). The bottom-right panel of Fig. ~\ref{cflls} suggests that the \ion{H}{i} covering fraction of sub-haloes is dependent on their specific star formation rate. We can explain this apparent dependence by noting that for sub-haloes $\vir$ refers to the tidal radius, which depends on the distance $d_{\mathrm{host}}$ to the center of the main halo.
Specifically, for sub-haloes located far from the host, the tidal radius approaches the definition of the virial radius of main haloes, whereas for those close to the host, the tidal radius is significantly smaller. Consequently, sub-haloes with $d_{\mathrm{host}} > 0.7 \, R_{\mathrm{vir, host}}$ exhibit trends of $\cov$ as a function of sSFR that are very similar to those of main haloes. While for sub-haloes closer to the host, the much smaller tidal radius results in median covering fractions that approach unity. The characteristic shape seen in the bottom-right panel of Fig. ~\ref{cflls} emerges from the combination of these two trends for all sub-haloes.

\subsection{Differential covering fraction of LLSs} \label{dfcov}

\begin{figure*}
    \centering
    \includegraphics[width=\linewidth]{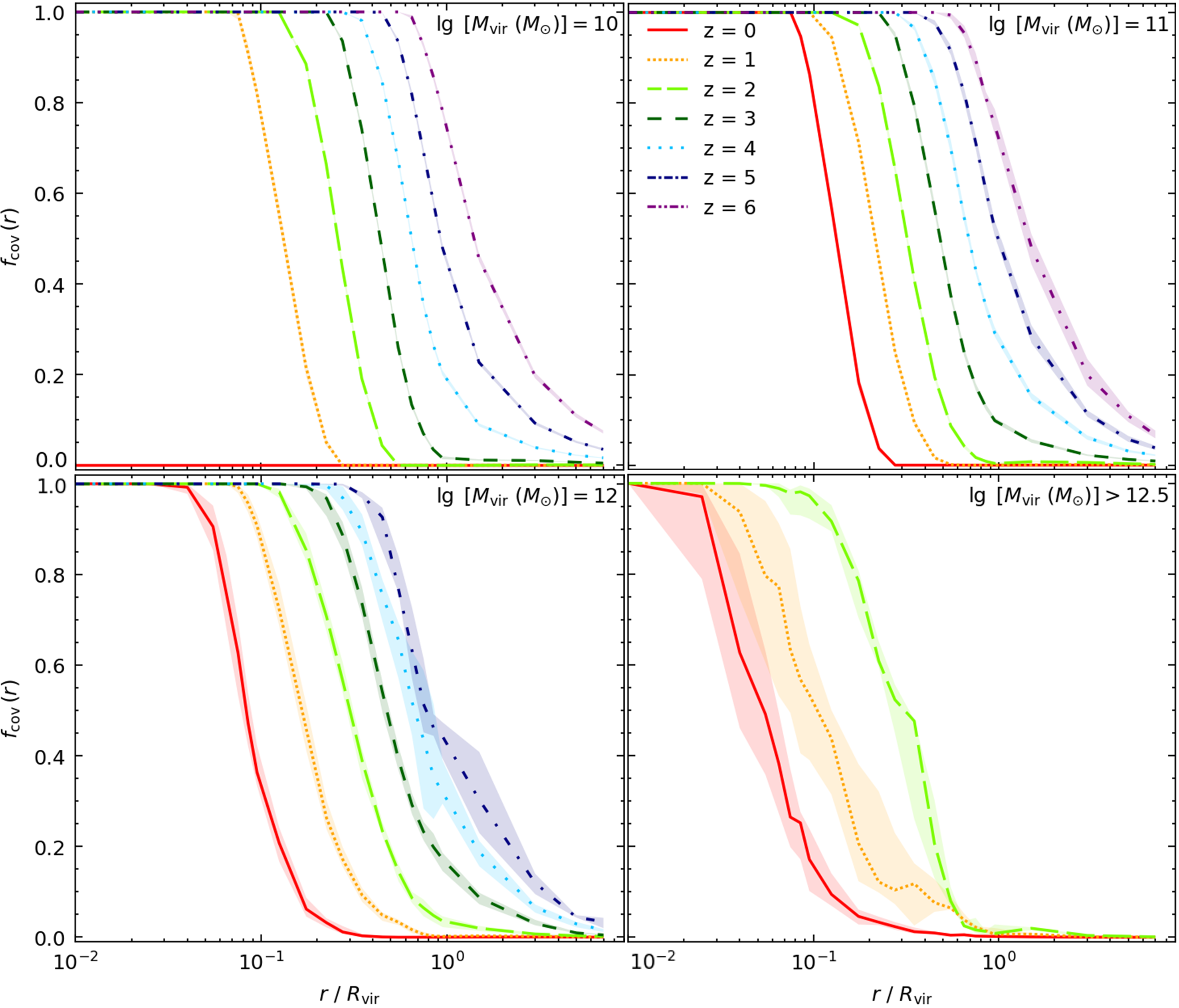}
    \caption{Differential covering fraction profiles of LLSs as a function of normalised impact parameter for four different mass bins. The different lines indicate the median $\covd$ at each impact parameter for different redshifts, and shaded areas show the 5th-95th percentile obtained from bootstrapping. The central mass of the bins is shown in the top right of each panel. All haloes with $M_{\mathrm{vir}} \pm 0.5 \, (0.3)$ dex are included in the $M_{\mathrm{vir}} = 10^{11}; 10^{12} \, (10^{10}) \, M_{\odot}$ bins, while only haloes with $ M_{\mathrm{vir}} > 10^{12.5} M_{\sun}$ are chosen for the bottom-right panel. The curves systematically shift to the left for decreasing redshift at all mass bins, indicating that the covering fraction of LLSs drops at all impact parameters for decreasing $z$.}
    \label{dflls_mass}
\end{figure*}
\begin{figure*}
	\includegraphics[width=\linewidth]{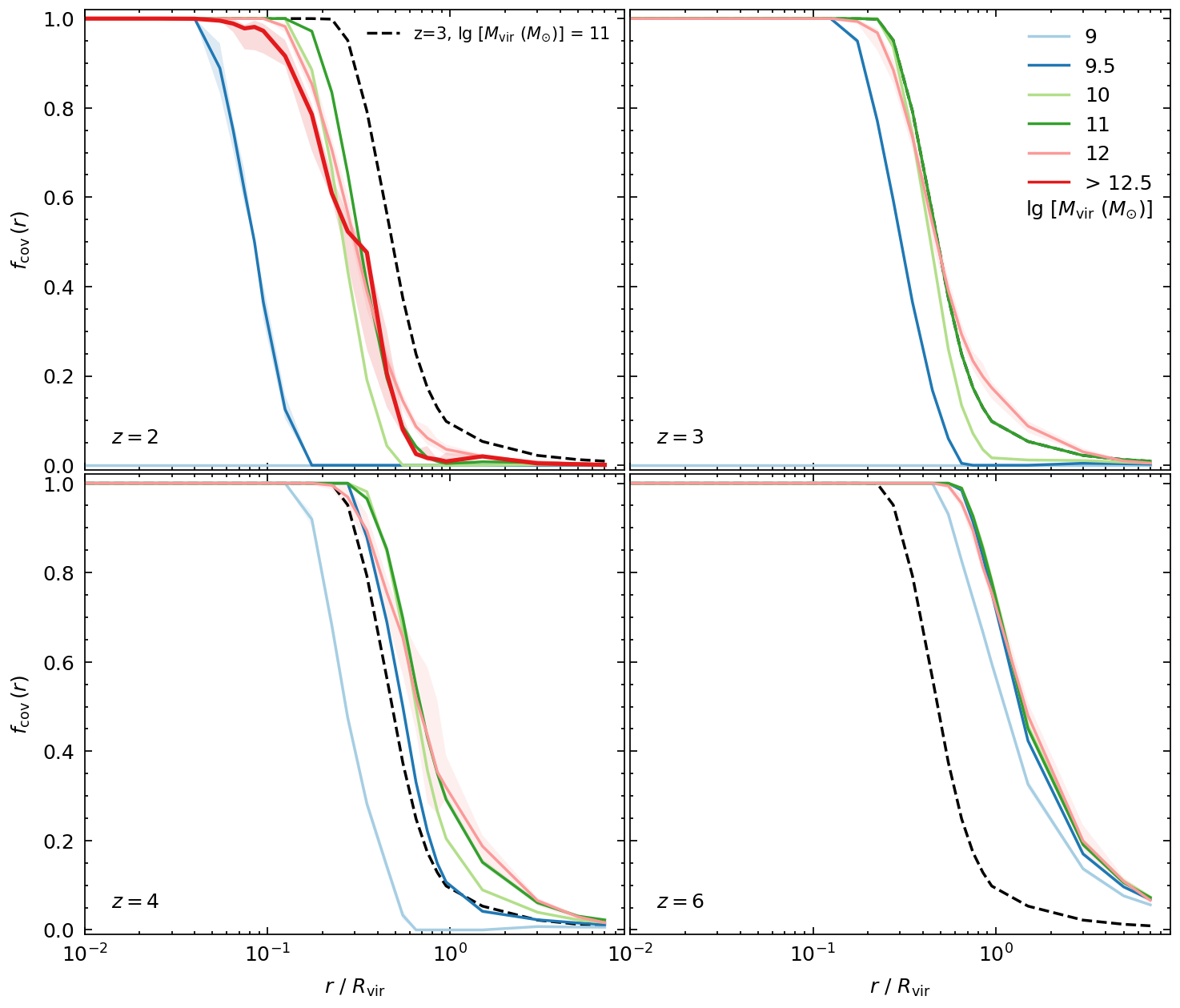}
    \caption{Differential covering fraction profiles of LLSs as a function of normalised impact parameters for redshifts $z \geq 2$. The solid lines indicate the median $\covd$ at each impact parameter for the different mass bins and the shaded areas show the 5th-95th percentile error on the median obtained from bootstrapping. The redshift is indicated at the bottom left of each panel. The reference profile of $M_{\mathrm{vir}} = 10^{11} M_{\sun}$ haloes at redshift $z=3$ is shown as a black-dashed line in every panel to aid in analysing the general trends. We see a strong increase in covering fraction at all impact parameters with increasing redshift. At these redshifts, we note that the differential covering fraction does not depend strongly on mass for haloes with $M_{\mathrm{vir}} = 10^{11} - 10^{12} M_{\sun}$, which hints at scale-invariance of the profiles. This does not hold for lower-mass ($M_{\mathrm{vir}} \leq 10^{10} M_{\sun}$) haloes, for which we note that the differential covering fraction profiles are generally less extended and evolve with halo mass.}
    \label{dflls_redshift}
\end{figure*}

In the previous section, we investigated the cumulative covering fractions of haloes. We now look at how the spatial distribution of LLSs around halo centers influences the covering fraction, by studying the differential covering fraction.
We show the profiles of $\covd$, disentangling the effects of mass and redshift, in Figures ~\ref{dflls_mass} \& ~\ref{dflls_redshift} respectively. We choose to depict the differential covering fraction as a function of normalised impact parameter $r \ / \ R_{\mathrm{vir}}$ guided by the previous study in \cite{2015MNRAS.452.2034R}. 

Fig. ~\ref{dflls_mass} shows the predicted median differential covering fraction for four mass bins, distinguishing between different redshifts. Each bin is centered at the indicated value and includes all haloes within $M_{\mathrm{vir}} \, \pm \, 0.5$ dex for $M_{\mathrm{vir}} = 10^{11}; 10^{12} M_{\odot}$ haloes and within $M_{\mathrm{vir}} \, \pm \, 0.3$ dex for $M_{\mathrm{vir}} = 10^{10} M_{\odot}$ haloes. For the last bin ($ M_{\mathrm{vir}} > 10^{12.5} M_{\sun}$), only haloes with mass above the threshold are chosen. We note that the profiles show higher values of differential covering fraction for all impact parameters with increasing redshift, for all mass bins. This result is consistent with the previous section and \cite{2023MNRAS.522.3831F}, namely that covering fractions of atomic hydrogen increase with increasing redshift. For instance, the pictures of the haloes show that \ion{H}{i} clouds extend far outside the virial radius of haloes for higher redshifts (see Fig. ~\ref{pic_redshift_same}, wherein we observe the filamentary structure around haloes to be more extended at higher redshifts). The differential covering fraction is thus expected to increase with redshift.
It is noted that the radial distribution of \ion{H}{i} around haloes has a sigmoidal shape, with some asymmetries that are more pronounced at higher redshift. As evident in the top-right panel of Fig. ~\ref{dflls_mass}, at redshift $z=0$, the median $\cov$ rapidly goes from 1 to 0 as the normalised impact parameter increases, with steep turning points. On the other hand, we see that for redshift $z=6$ the median $\cov$ steeply decreases from 1, but goes down more slowly towards 0 for large impact parameters away from the center.

In Fig. ~\ref{dflls_redshift} we show the predicted differential covering fraction for individual redshifts $z=2, 3, 4$ and 6, and classify the haloes in each panel by mass bins as in Fig. ~\ref{dflls_mass}. We include two lower-mass bins for this analysis, namely all haloes within $M_{\mathrm{vir}} \pm \, 0.1$ dex for $M_{\mathrm{vir}} = 10^9 \ M_{\sun}$ haloes and within $M_{\mathrm{vir}} \, \pm \, 0.2$ dex for $M_{\mathrm{vir}} = 10^{9.5} \ M_{\sun}$ haloes. At these redshifts, we find that the profiles of haloes with $M_{\mathrm{vir}} \simeq 10^{11}-10^{12} M_{\sun}$ take roughly the same values at all impact parameters. The curves are almost superimposed, hinting at the existence of some characteristic length scale similar to the virial radius for this mass of haloes \citep{2014MNRAS.438..529R, 2015MNRAS.452.2034R}. This explains the weaker dependence (i.e. flattening) of $\cov$ noted in haloes with $M_{\mathrm{vir}} \gtrsim 10^{11} M_{\sun}$ (Fig. ~\ref{cflls}). Although the specific geometry and physical scale of individual haloes are different, we find that the gas is distributed to a similar relative extent away from their center. This hints at scale invariance, which is studied in the following subsection. We note that a similar trend is found for redshifts $z=0$ and 1 (not shown in Fig. ~\ref{dflls_redshift}), although with lesser agreement.

The scale-invariance observed in more massive haloes does not seem to hold for $M_{\mathrm{vir}} \leq 10^{10} M_{\sun}$ haloes. This explains the mass-dependence of the cumulative covering fraction of $M_{\mathrm{vir}} \lesssim 10^{11} M_{\sun}$ haloes (see Fig. ~\ref{cflls}). Indeed, the LLS differential covering fraction profiles of $10^{9}-10^{10} M_{\sun}$ haloes are shifted to smaller radii, indicating that neutral hydrogen is less extended in lower mass haloes and hence the cumulative covering fraction is also smaller. As redshift increases, we find that the radial profiles of $10^{9}-10^{10} M_{\sun}$ haloes evolve from being zero at all radii to gradually shifting to higher radii, until they nearly converge with those of massive haloes at $z=6$. The evolution of the cumulative covering fraction of LLSs with halo mass is thus explained by the evolving radial concentration of \ion{H}{i} with halo mass. A further investigation reveals that the \ion{H}{i} column density in lower mass haloes is distributed similarly to that in more massive haloes in the outskirts ($\gtrsim \vir$) but is significantly lower in the central regions, see Fig. ~\ref{fig:nhi_profile} in the appendix. Consequently, the density threshold of Lyman Limit Systems is reached at smaller radii in lower mass haloes and the resulting differential covering fraction profiles are thus less extended. This suggests that gas which is located in the center of more massive haloes is expulsed from the halo and redistributed via heating by the UV background, stellar feedback and tidal interactions in lower-mass haloes \citep{2020MNRAS.498.4745J, 2023MNRAS.524.5391A, 2024ApJ...960...55Z}.

There is tentative evidence that the scale-invariance is also not exhibited in the most massive haloes ($M_{\mathrm{vir}} \geq 10^{12.5} M_{\sun}$), particularly at redshifts $z=0,1$ (not shown here). It was found in other studies that, at this mass, the cooling time of gas in the inner parts of massive haloes is long, such that there is a low fraction of neutral gas there, which could explain the lower differential covering fraction of \ion{H}{i} \citep{2021MNRAS.507.2869S}. 
However, there are only a few objects in our simulation that reach these masses at redshifts $z=0, 1$ and 2 (respectively: 12, 5 and 1), and a more robust sample is needed to draw conclusions.

\subsection{Fitting function for differential covering fraction}\label{sec:fitfct}

\begin{table}
\centering
\begin{tabular}{ c || c | c | c | c | c | }
 \hline
 Halo mass [$M_{\sun}$] & Parameter & $a$ & $b$ & $c$ & $d$ \\
 \hline
 \multirow{4}{*}{$M_{\mathrm{vir}} = 10^{11}$} & $A$ & $-0.0065$ & $0.092$ & $-0.153$ & $0.012$ \\
 & $B$ & $0.0017$ & $-0.016$ & $0.026$ & $0.998$ \\
 & $L_z$ & $0.0008$ & $0.005$ & $0.084$ & $0.13$ \\

 & $\alpha$ & $-0.06$ & $0.57$ & $-1.58$ & $5.86$ \\
 \hline
 \multirow{4}{*}{$M_{\mathrm{vir}} = 10^{12}$} & $A$ & $-0.0013$ & $0.054$ & $-0.061$ & $0.034$ \\

 & $B$ & $0.0003$ & $-0.0070$ & $0.0086$ & $0.989$ \\
                                                       
 & $L_z$ & $-0.007$ & $0.045$ & $0.041$ & $0.085$ \\

 & $\alpha$ & $0.099$ & $-0.39$ & $-0.017$ & $4.33$ \\
 \hline
\end{tabular}
\caption{Best-fit values for the free parameters of the differential covering fraction fitting function for LLSs of haloes with $M_{\mathrm{vir}} = 10^{11} M_{\sun}$ or $10^{12} M_{\sun}$.}
\label{tablefit}
\end{table}

\begin{figure*}
    \centering
    \includegraphics[width=\linewidth]{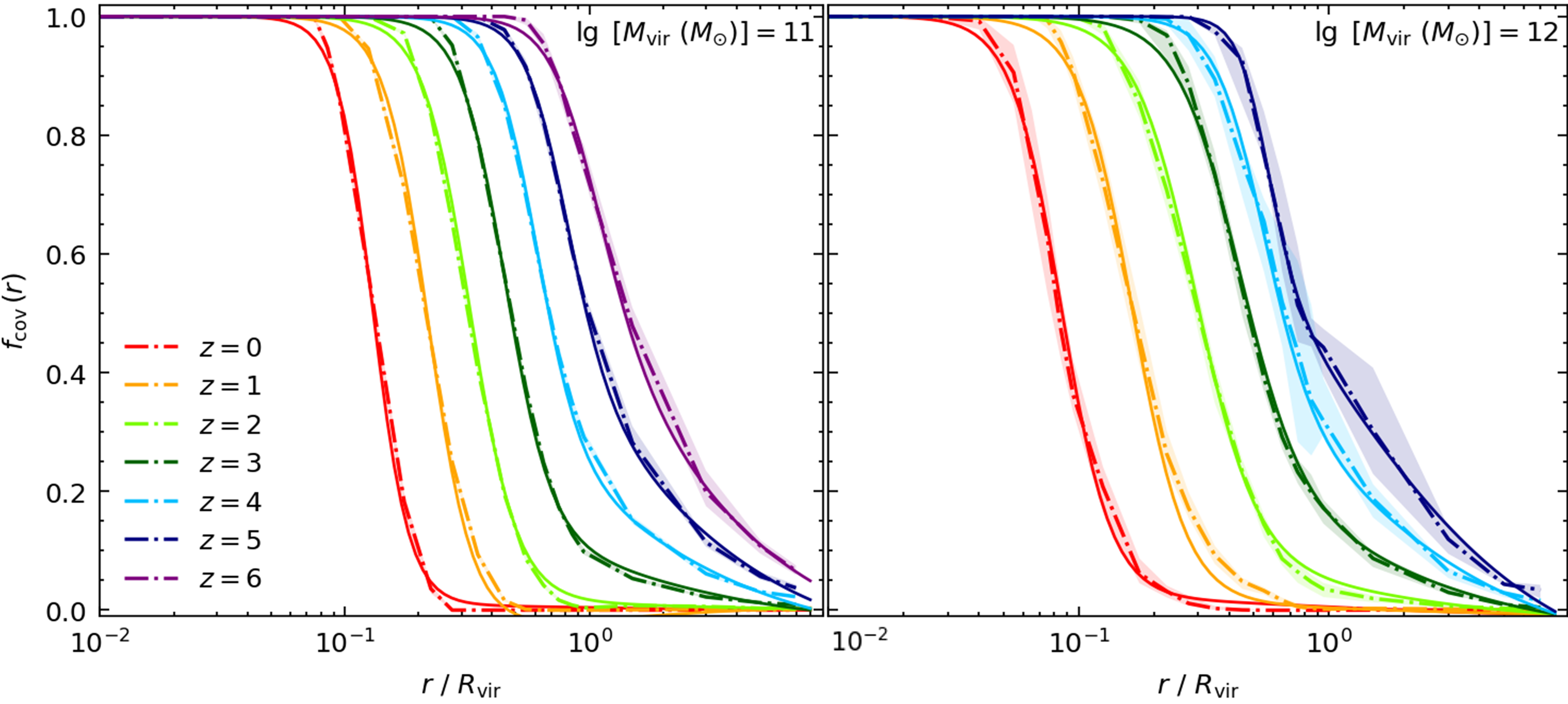}
    \caption{Comparison of differential covering fraction profiles obtained from the simulations (dashed lines) and the fitting function of equation \eqref{fitfunc} (solid lines). The different colours distinguish between different redshifts, and the mass bin of the haloes considered is shown at the top right of each panel. We note that the fit performs very well in recovering the features of the radial profiles and that we can attribute some physical length scale to certain parameters (see the main text for details).
    }
    \label{fit}
\end{figure*}

Our study of the differential covering fraction of LLSs showed that haloes of a certain mass range share very similar, possibly scale-invariant, profiles. This can be further investigated by way of a generalised fitting function for the radial profiles of atomic hydrogen around haloes. Let us denote the normalised impact parameter as $x = r \  / \ \vir$. The differential covering fraction of LLSs around haloes with $M_{\mathrm{vir}} \simeq 10^{11}-10^{12} M_{\sun}$ at a given redshift $z$ can be fitted via:
\begin{equation}\label{fitfunc}
    f_{\mathrm{cov}}(x, z) = 1 + \left( \frac{A(z)}{1+x} - 1\right) \cdot \frac{1}{B(z) + \left(L_z(z) /
    x\right)^{\alpha(z)}}
\end{equation}
where the four free parameters $A, B, L_z$ and $\alpha$ are fitted by way of a 3-rd degree polynomial in redshift $z$. This fitting function is a revision of a similar characterization of differential covering fraction profiles of LLSs proposed in \cite{2015MNRAS.452.2034R} (see Eq. (5) therein). It satisfies important properties that a physical distribution should respect in the appropriate limits. In particular, it approaches unity in the two limits $x\to 0$ and $z \to \infty$, and approaches some asymptotic value $1 - 1/B$ at large impact parameters, which depends on redshift.

The parameters determining our empirical fit are physically meaningful. For instance, $L_z$ can be interpreted as some typical projected distance between galaxies and \ion{H}{i} absorbers \citep[similarly to][]{2015MNRAS.452.2034R}. $x \sim L_{z}$ corresponds roughly to where $\covd\sim 0.5$, and hence the parameter $L_z$ can be used to estimate the distance between haloes-absorbers. $A$ dictates the first turning point where $\covd$ decreases from 1; $B$ determines the asymptotic value of the differential covering fraction of haloes at large impact parameters; $\alpha$ roughly describes the slope of $\covd$ (i.e., how quickly it goes from 1 to 0 as a function of $x$). All these parameters $P$ are described via $P(z) = a\cdot z^3 + b\cdot z^2 + b\cdot z + d$. The values for the best-fit coefficients for each parameter used in equation \eqref{fitfunc} are summarised in Table ~\ref{tablefit} and shown in Figure ~\ref{fit}.

At all redshifts, the fitting function reproduces accurately the characteristic behaviour we described in section ~\ref{dfcov}. We remark that it is not as precise at redshifts $z=0$ and 1, but this is expected as it corresponds to the redshifts for which the actual profiles are the least superimposed. The fitting function highlights the asymmetries of the differential covering fraction profiles observed in the previous section. Using the values listed in Table ~\ref{tablefit}, we find that at $z \sim 3$ the expected projected distance between LLSs and their host haloes should be around $L_z \approx 0.42-0.45 \ R_{\mathrm{vir}}$ for $M_{\mathrm{vir}} = 10^{12}-10^{11} M_{\odot}$ haloes respectively. These values are smaller than findings from the OWLS simulations \citep{2014MNRAS.438..529R} and from EAGLE \citep{2015MNRAS.452.2034R}, wherein it was found that such distance $L_z \approx R_{\mathrm{vir}}$. This suggests that \ion{H}{i} gas in FIREbox is concentrated closer to the center of haloes than in the EAGLE simulations.

\section{DISCUSSION} \label{discussion}

\subsection{Cumulative covering fraction}

The most recent and statistically significant constraints from observations of the \ion{H}{i} covering fraction of Lyman Limit Systems come from the Keck Baryonic Structure Survey \citep{2012ApJ...750...67R, 2014ApJ...795..165S, 2017ApJ...836..164S} and the Quasars Probing Quasars project \citep[QPQ; see][and references therein]{2018ApJS..236...44F}. On the one hand, \cite{2012ApJ...750...67R} report a \ion{H}{i} covering fraction $\cov = 0.30 \pm 0.14$ around Lyman Break Galaxies (LBGs) at $z \sim 2 - 2.5$, residing in haloes with $M_{\mathrm{vir}} \approx 10^{12} M_{\sun}$.  On the other hand, \cite{2013ApJ...762L..19P} predict a covering fraction $\cov = 0.64^{+0.06}_{-0.07}$ around $z \sim 2 - 2.5$ quasars (QSOs) residing in haloes with characteristic halo mass of $M_{\mathrm{vir}} \approx 10^{12.5} M_{\sun}$ \citep{2012MNRAS.424..933W}. These results have historically been a challenge to reproduce in simulations, and different suites and physical implementations lead to varying predictions.

The most recent numerical works carried out on this topic \citep[]{2014ApJ...780...74F, 2015MNRAS.452.2034R, 2015MNRAS.449..987F, 2016MNRAS.461L..32F, 2015MNRAS.453..899M, 2017MNRAS.468.1893M, 2017MNRAS.464.2796G, 2019MNRAS.483.4040S} have all been able to broadly reproduce covering fractions found around LBGs by \citet{2012ApJ...750...67R}, while using a vast range of numerical solvers and sub-grid physics. The high covering fractions observed around QSOs by \citet{2013ApJ...762L..19P} have posed a greater challenge to reproduce \citep[see e.g.,][]{2014ApJ...780...74F, 2015MNRAS.449..987F}. Nonetheless, further work conducted by \citet{2016MNRAS.461L..32F} was able to replicate values for the QSOs using higher resolution zoom-ins and the same (stellar feedback driven) physics, arguing that the high resolution enabling more finely-resolved stellar feedback from satellites is a key ingredient in matching the observations. Additionally, \citet{2015MNRAS.452.2034R} have succeeded in matching both observations using the EAGLE suite of cosmological simulations \citep{2015MNRAS.450.1937C, 2015MNRAS.446..521S}, via implementation of both stellar and AGN feedback at a lower numerical resolution.

We now discuss the meaning of our results and compare them with those of the previous works described above.
We found that the covering fraction of LLSs in FIREbox haloes increases with increasing redshift, and is roughly independent of mass in massive ($M_{\mathrm{vir}} \gtrsim 10^{11}-10^{12} M_{\sun}$) haloes. These conclusions are largely in agreement with other simulations \citep[see][]{2014ApJ...780...74F, 2015MNRAS.449..987F, 2015MNRAS.452.2034R, 2017MNRAS.464.2796G, 2021MNRAS.507.2869S}. 
Our values of \ion{H}{i} covering fraction for the very-massive ($M_{\mathrm{vir}} \gtrsim 10^{12.5} M_{\sun}$) haloes are lower than found in other cosmological-size suites as in \citet{2015MNRAS.452.2034R}. They report that the covering fraction in $M_{\mathrm{vir}} \approx 10^{12.5} M_{\sun}$ haloes is roughly $\cov \approx 0.27$ at $z=2$, whereas \fb{} results at the same redshift are about $\cov \sim 0.15-0.2$. This overall trend is the same at redshifts $z=1, 3$ and 4.
We note, however, that there are significantly fewer of these very-massive haloes in our simulation than in \citet{2015MNRAS.452.2034R}. Specifically, there is only one very-massive halo in FIREbox at $z=2$, whereas EAGLE has 39 very-massive haloes at redshift $z=3$ and 116 at redshift $z=2$. \\

\begin{figure}
    \centering
    \includegraphics[width=\linewidth]{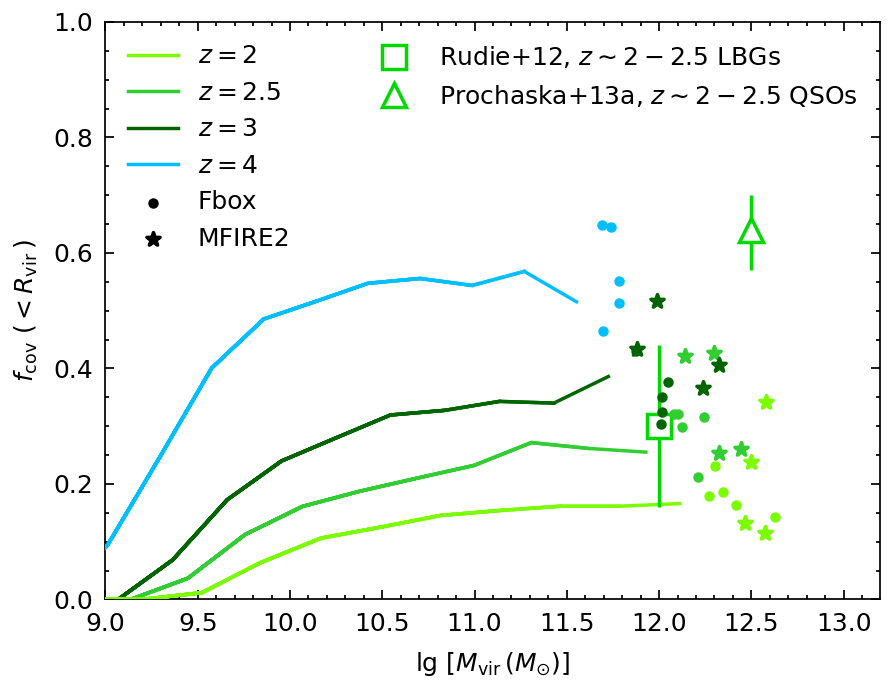}
    \caption{Cumulative covering fraction of LLSs in FIRE simulations and in observations at redshifts $z=2-4$. \textit{Solid lines}: Median cumulative covering fraction from FIREbox for different mass bins. \textit{Coloured circles}: Average (over three orthogonal projections) of the cumulative covering fraction of the 5 most massive haloes at each redshift in FIREbox. \textit{Coloured stars}: Average (over three orthogonal projections) of the cumulative covering fraction of the 4 most massive haloes at each redshift in the MassiveFIRE (FIRE-2) runs (see main text for details). The lines and points are colour-coded by redshift as per the rest of the paper and shown in the top-left of the figure. 
    We compare the simulations with observations, indicated by a square \citep{2012ApJ...750...67R} and a triangle \citep{2013ApJ...762L..19P}.
    We note that the covering fractions from the LBGs sample match well with the expectations for $M_{\mathrm{vir}} \approx 10^{12} M_{\odot}$ FIREbox haloes at redshift $z=2-2.5$. However, the observed covering fractions from QSOs are neither reproduced by FIREbox nor MassiveFIRE (FIRE-2) simulations.} 
    \label{obs_tot}
\end{figure}

We proceed to compare our results with observations. Given the above discussions, we complement our FIREbox results for the very-massive haloes by adding 4 ‘zoom-in' simulations of $M_{\mathrm{vir}} \, (z=2) \sim 10^{12.5} M_{\sun}$ haloes run with the same FIRE-2 physics (see section ~\ref{method} for details). The results are shown in Figure ~\ref{obs_tot}. We highlight the LLS covering fractions predicted by FIREbox for redshifts $z=2, 2.5, 3, 4$, the results for our zoom-ins, and the data obtained by \citet{2012ApJ...750...67R, 2013ApJ...762L..19P}. Our results for redshifts $z=2, 2.5$ are well within the confidence interval of the LBGs observations, and we predict that the sample used to obtain these covering fractions is matched similarly well by $10^{12} M_{\sun}$ haloes at redshifts $z=2, 2.5$. This conclusion is consistent with other numerical works of similar scope \citep[see e.g.][]{2015MNRAS.449..987F, 2016MNRAS.461L..32F, 2015MNRAS.452.2034R}. 

Our FIRE-2 simulations do not reproduce the high covering fraction observed in QSOs by \citet{2013ApJ...762L..19P}. The zoom-in haloes presented here tend to have a larger range of covering fractions and span varied accretion histories. One halo, in particular, shows a higher average covering fraction of $\cov \approx 0.35$ at $z=2$, but the zoom-ins do not constitute any major improvement against the QSOs observations. This is to be contrasted with results from \citet{2016MNRAS.461L..32F}, wherein the large \ion{H}{i} covering fraction in very-massive haloes was attributed to the enhanced resolution of low-mass satellite galaxies and their associated winds interacting with filaments of cosmic origin in the zoom-ins.

Our investigation reveals that, on average, the covering fraction of our haloes is lower than the mean covering fraction in the sample introduced in \citet{2016MNRAS.461L..32F}. 
Several factors contribute to this disparity. On the one hand, they analyzed a more extensive ensemble of 15 haloes to our limited sample of 4, granting them more robustness in analyzing mean values of covering fraction. Specifically, at redshift $z=2$ and for $M_{\mathrm{vir}} \approx 10^{12.5} M_{\odot}$ haloes, their analysis revealed a broad halo-to-halo scatter of $\cov \in [0.3, 0.6]$ while the average exceeded $\cov \gtrsim 0.5$, bringing them considerably closer to the observed value for QSOs. It is hence plausible that the values we obtain from our FIRE-2 simulations are on the low end of a distribution, and that selecting more haloes to simulate in zoom-ins might raise the average covering fraction at the high-mass end. Furthermore, their analysis includes very-massive haloes at higher redshift, with $M_{\mathrm{vir}} \gtrsim 10^{12.5} M_{\odot}$ at $z=2.5$. This is significant because of the notable surge in the typical covering fraction from $z=2$ to $z=2.5$, naturally producing a higher mean covering fraction when including such haloes, which we have not done in this work. \\

Our analysis of the cumulative covering fraction in $M_{\mathrm{vir}} < 10^{12.5} M_{\odot}$ haloes demonstrates good agreement with observational data for $M_{\mathrm{vir}} = 10^{12} M_{\odot}$ LBGs. This work robustly extends the statistics down to very-low masses of $M_{\mathrm{vir}} \sim 10^{8} M_{\odot}$, unveiling a complex halo-mass dependence of the \ion{H}{i} covering fraction. Given the very low number of $M_{\mathrm{vir}} \gtrsim 10^{12.5} M_{\odot}$ haloes in our sample from \fb{} supplemented with 4 MassiveFIRE (FIRE-2) haloes, we cannot robustly compare our covering fractions with the QSOs sample and leave this for future work.

\subsection{Differential covering fraction}

\begin{figure}
    \centering
    \includegraphics[width=\linewidth]{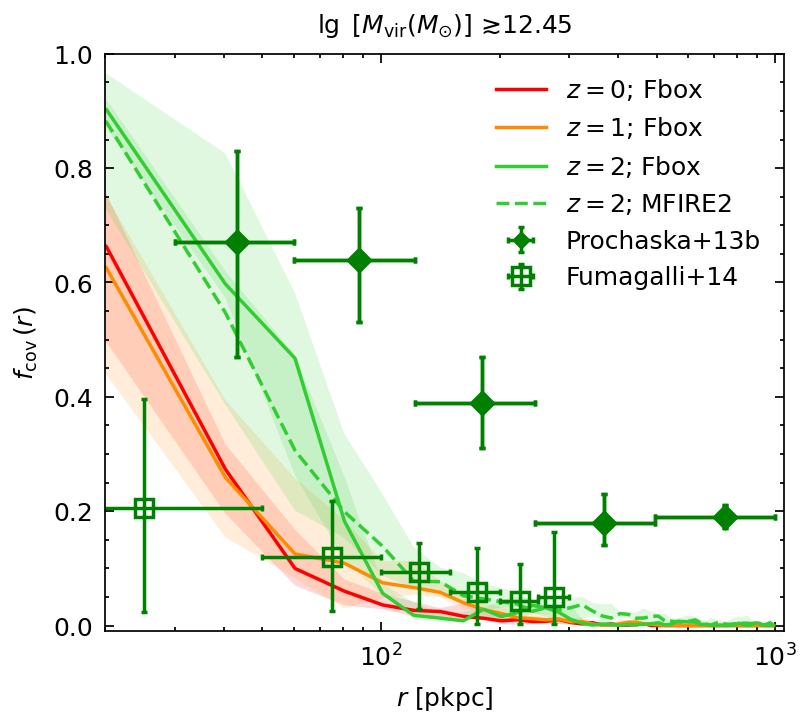}
    \caption{Simulated and observed differential \ion{H}{i} covering fraction for redshift $z\sim2$. \textit{Solid lines}: Median differential covering fraction as a function of physical projected impact parameter away from the center of $M_{\mathrm{vir}} \gtrsim 10^{12.5} M_{\sun}$ haloes in FIREbox (1 halo at $z=2$). \textit{Dashed line}: Median differential covering fraction as a function of physical projected impact parameter away from the center of the four most massive haloes in the MassiveFIRE (FIRE-2) zoom-ins (4 haloes at $z=2$). In both cases, the shaded areas display the 5th-95th percentile error on the median obtained from bootstrapping. Note that there are no haloes of this mass at redshift $z=2.5$ in either the FIREbox or MassiveFIRE (FIRE-2) simulations. \textit{Data points}: the diamond data points show the observations of \citet{2013ApJ...776..136P} for a sample of quasars at $z \sim 2-2.5$. The square data points show the simulations of \citet{2014ApJ...780...74F} for five $M_{\mathrm{vir}} \gtrsim 10^{12.5} M_{\sun}$ galaxies at $z=2$. In both, the scatter in the abscissa indicates the lowest and highest impact parameter of each bin for which the value of $\covd$ is calculated. We find good agreement with observations close to the halo centers ($\leq 50$ kpc) but systematically underestimate the covering fraction at large impact parameters ($\geq 100$ kpc).}
    \label{dflls_obs}
\end{figure}

We continue our comparisons with previous works, turning now to the radial profiles of covering fraction of LLSs. \cite{2015MNRAS.452.2034R} were able to reconcile observations around quasars by comparing results in physical units rather than fractions of the virial radius. We follow this convention in this section and discuss its implications in the text.

In Figure ~\ref{dflls_obs}, we present our results for the differential covering fraction of very-massive ($M_{\mathrm{vir}} \gtrsim 10^{12.45} M_{\odot}$) haloes in both FIREbox and MassiveFIRE (FIRE-2), with data points for previous simulations and observations. The simulations by \cite{2014ApJ...780...74F} reported very low covering fractions of LLSs which did not match the observations at any impact parameter. \cite{2015MNRAS.452.2034R} (not shown in Fig. ~\ref{dflls_obs}) report excellent agreement between observed and simulated covering fractions of LLSs with the EAGLE simulations, at all impact parameters. Both FIREbox and MassiveFIRE (FIRE-2) results agree with observations of radial profiles of \ion{H}{i} in the vicinity of very-massive haloes, but systematically underestimate the covering fraction further away from their center. In particular, we predict that the covering fraction of LLSs tends to 0 as $r$ increases, effectively becoming null for gas extending beyond $\sim 800$ kpc from the center, whereas the quasar sample from \cite{2013ApJ...762L..19P} suggests that it stagnates at a non-zero value of $\covd \approx 0.2$.

FIREbox underestimates radial profiles of the observed covering fraction of LLSs, particularly in the outer regions of haloes. This particular outcome is not an exception: it was noted in most numerical works which could not reproduce the QSOs values \citep[e.g.,][]{2015MNRAS.453..899M, 2017MNRAS.468.1893M, 2017MNRAS.464.2796G, 2019MNRAS.483.4040S}, and in related observing campaigns using the QPQ data \citep{2015ApJ...808...38R}. This result can be expected for FIREbox, which slightly underestimates the overall covering fraction of LLSs around very-massive haloes, seemingly due to a smaller amount of \ion{H}{i} found at large radii away from the center of haloes. 

\begin{figure*}
	\includegraphics[width=\linewidth]{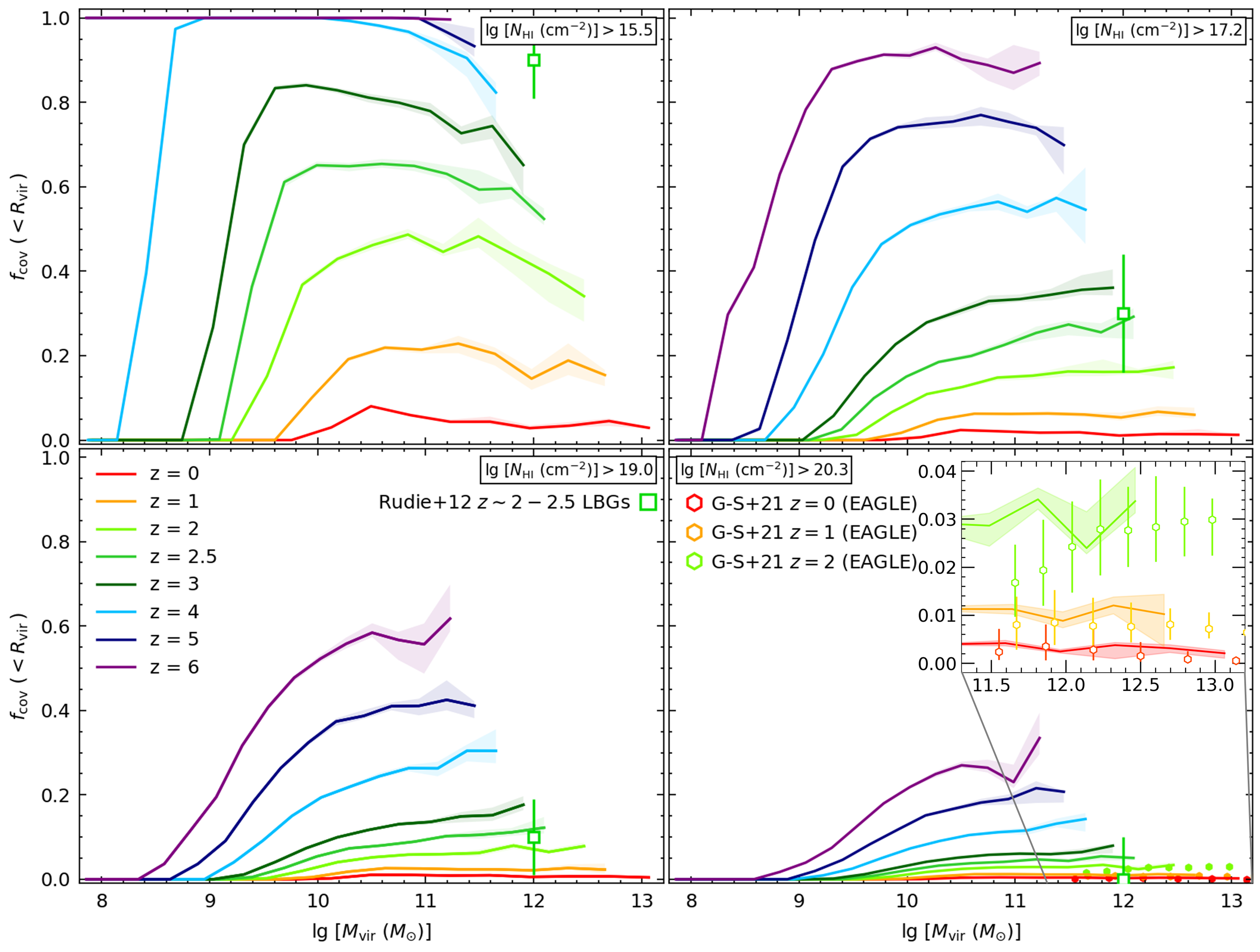}
    \caption{\ion{H}{i} cumulative covering fraction for \fb{} main haloes within impact parameter $\vir$ as a function of halo mass $M_{\mathrm{vir}}$ for diverse density cuts. From top-left to bottom-right, we show $\cov$ for \ion{H}{i} density cuts of $N_{\ion{H}{i}} > 15.5$ cm$^{-2}$ (sLLSs), $N_{\ion{H}{i}} > 17.2$ cm$^{-2}$ (LLSs; shown here for reference), $N_{\ion{H}{i}} > 19.0$ cm$^{-2}$ (sDLAs) and $N_{\ion{H}{i}} > 20.3$ cm$^{-2}$ (DLAs). Different colours are used to distinguish between redshift from $z=0$ to $z=6$ as per the legend. Solid lines and shaded areas respectively show the median cumulative covering fraction and the 5th-95th percentile error on the median obtained from bootstrapping. All haloes with $M_{\mathrm{vir}} \geq 10^{7.75} M_{\sun}$ identified in the simulation were used to create these plots. Square data points correspond to measurements around $z\sim2, 2.5$ LBGs at the corresponding density cuts from \citet{2012ApJ...750...67R} and hexagonal data points indicate the median and 16th-84th percentile of the covering fraction for a sample of DLAs from the EAGLE simulations reported in \citet{2021MNRAS.501.4396G}. We find good agreement with the covering fractions from the LBGs sample for LLSs, sDLAs and DLAs. For the sLLS threshold ($N_{\ion{H}{i}} > 15.5$ cm$^{-2}$), we find that \fb{} underestimates the cumulative covering fraction by $\sim$30\%. FIREbox shows good agreement with the EAGLE simulations at low redshifts and extends the relation down to low masses and to higher redshifts.}
    \label{cfnhi}
\end{figure*}

In the EAGLE simulations, \citet{2015MNRAS.452.2034R} find agreement with observations of differential covering fraction by \citet{2013ApJ...776..136P} for $M_{\mathrm{vir}} \gtrsim 10^{12.5} M_{\odot}$ haloes. We stress however that \citet{2015MNRAS.452.2034R} cannot reproduce the observed cumulative covering fraction $\cov = 0.64^{+0.06}_{-0.07}$, underestimating it by a factor $>2$. The agreement here comes from considering the observational biases present in the QPQ sample (for details, see discussions in \citet{2015MNRAS.452.2034R} and \citet{2016MNRAS.461L..32F}). Essentially, they argue that \citet{2013ApJ...776..136P} overestimate the covering fraction because of using a fixed virial radius typical of $M_{\mathrm{vir}} \approx 10^{12.5} M_{\sun}$ haloes, whereas they probe quasars of higher mass than predicted, closer to $M_{\mathrm{vir}} \gtrsim 10^{13} M_{\sun}$. They further argue that most of the sight lines at high impact parameters actually come from objects with redshift $z > 2$. Both effects essentially lead to an overestimation of the cumulative covering fraction in the QPQ sample, particularly at high impact parameters away from the center, and \citet{2015MNRAS.452.2034R} find agreement with observations by correcting for these effects. Given the absence of haloes exceeding $M_{\mathrm{vir}} \gtrsim 10^{13} M_{\sun}$ at redshifts $z=2-2.5$ in our FIREbox and MassiveFIRE (FIRE-2) simulations, we were unable to employ this method to our results, and hence cannot conclude on its effectiveness in recovering the observed radial profiles of covering fraction.

Our discussion underscores the lack of consensus and challenges in comparing observations and simulations of \ion{H}{i} covering fraction, as well as the large range of predictions produced by different models. 

\subsection{Cumulative covering fraction for additional $N_{\ion{H}{i}}$ cuts (sLLS, LLS, sDLA, DLA)}

To offer a more complete overview of strong \ion{H}{i} absorbers in FIRE, we extend our analysis of the covering fraction of Lyman Limit Systems by providing the \ion{H}{i} covering fraction of haloes in \fb{} at additional density cuts. The methods for computing these covering fractions are identical to that introduced in section ~\ref{method}. Figure ~\ref{cfnhi} shows the covering fraction of atomic hydrogen $\cov$ of main haloes in \fb{} as a function of halo virial mass, for density cuts corresponding to the class of absorbers known as sub-Lyman Limit Systems (sLLSs; $N_{\ion{H}{i}} > 15.5$ cm$^{-2}$), sub-Damped Lyman-$\alpha$ Absorbers (sDLAs; $N_{\ion{H}{i}} > 19.0$ cm$^{-2}$) and Damped Lyman-$\alpha$ Absorbers (DLAs; $N_{\ion{H}{i}} > 20.3$ cm$^{-2}$). We also compare with measurements by \citet{2012ApJ...750...67R} of the \ion{H}{i} covering fraction at these density cuts in each panel and show results for the cumulative covering fraction of DLAs in the EAGLE simulations from \citet{2021MNRAS.501.4396G}. 

The covering fractions measured by \citet{2012ApJ...750...67R} for sDLAs and DLAs are in good agreement with results from \fb{}, as was the case for LLSs (Fig. ~\ref{obs_tot}). For the lower column density threshold of $N_{\ion{H}{i}} > 15.5$ cm$^{-2}$, the simulations slightly underestimate the cumulative covering fractions of $M_{\mathrm{vir}} = 10^{12} M_{\sun}$ LBGs at redshifts $z\sim2-2.5$. In the zoomed view of the bottom-left panel of Fig. ~\ref{cfnhi}, we show that the covering fractions of DLAs in \fb{} are similar to those of EAGLE massive haloes reported by \citet{2021MNRAS.501.4396G}. Our analysis robustly extends the redshift and mass dependence for these systems, offering greater possibilities to compare results from future simulations and observations.

\section{SUMMARY AND CONCLUSIONS}\label{conclusions}

In this work, we have used simulations run with the FIRE-2 physics prescription \citep{2018MNRAS.480..800H} at high numerical resolution ($m_{\mathrm{b}} = 3.6-6.24 \times 10^4 M_{\sun}$) to comprehensively investigate the atomic hydrogen covering fraction of Lyman Limit Systems in haloes spanning the mass range from $M_{\mathrm{vir}} \sim 10^{12.5} M_{\sun}$ down to very-low mass haloes with $M_{\mathrm{vir}} \sim 10^{8} M_{\sun}$, across redshifts $z=0-6$. Our analysis includes haloes from the (22.1 cMpc)$^{3}$ cosmological volume simulation \fb{} \citep{2023MNRAS.522.3831F}, which currently constitutes the highest-resolution cosmological volume simulation with the largest dynamical range of its kind, supplemented with zoom-ins of four massive haloes from the MassiveFIRE (FIRE-2) sample \citep{2016MNRAS.458L..14F, 2017MNRAS.470.1050F, 2017MNRAS.470.4698A}. This comprehensive approach allows for a more rigorous comparison with both observations and prior numerical investigations on the subject.
 
Our main results can be summarized as follows:
\begin{itemize}
    \item The \ion{H}{i} cumulative covering fraction of LLSs in FIREbox exhibits a pronounced dependence on redshift, showing significant increase at all halo masses from redshift $z=0$ to $z=6$ (Fig. ~\ref{cflls}). The complex halo mass dependence of $\cov$ can be divided into two regimes. Notably, the high resolution of our study enables an investigation of the dependence for low-mass haloes, revealing that the cumulative covering fraction steeply increases from zero to some maximal value, which increases with increasing redshift, from $M_{\mathrm{vir}} = 10^{7.75} M_{\odot}$ to a threshold mass which decreases with increasing redshift. For instance, at $z=0$, this threshold resides at $\ M_{\mathrm{vir}} \sim 10^{10.5} M_{\odot}$ while at $z=6$, it is approximately $\ M_{\mathrm{vir}} \sim 10^{9.2} M_{\odot}$. Beyond this threshold, the covering fraction plateaus and remains nearly independent of mass for higher-mass haloes at all redshifts.
    \item The \ion{H}{i} differential covering fraction of LLSs in FIREbox is also highly dependent on redshift, showing a similar increase at all halo masses from redshift $z=0$ to $z=6$, see Fig. ~\ref{dflls_mass}. The radial profiles $\covd$ as a function of projected impact parameter $r$ from the center are found to resemble an inverse-sigmoid. It takes the value of $\covd = 1$ for inner impact parameters close to the center, before steeply decreasing toward 0 further away from the center.
    \item The turning points in the radial profiles happen at lower impact parameters with decreasing redshift, indicating that a greater fraction of \ion{H}{i} is found closer to the center of haloes with decreasing redshift (Fig. ~\ref{dflls_redshift}). In particular, we note that almost all of the \ion{H}{i} is found within the virial radius of haloes at lower redshifts ($z \lesssim 2$), in agreement with \citet{2018ApJ...866..135V, 2023MNRAS.522.3831F}.
    \item The radial profiles of strong \ion{H}{i} absorbers in massive ($M_{\mathrm{vir}} \simeq 10^{11}-10^{12} M_{\odot}$) haloes in FIREbox are very similar to each other and show scale-invariance, see Fig. ~\ref{dflls_redshift}. 
    \item Lower mass ($M_{\mathrm{vir}} \leq 10^{10} M_{\odot}$) haloes have differential covering fraction profiles with a similar shape but are significantly less extended than more massive haloes. The profiles get more extended with increasing mass and redshift until they overlap at all halo masses. This explains the evolution of the cumulative covering fraction with mass and redshift (Fig. ~\ref{cflls} and Fig. ~\ref{dflls_redshift}).
    \item We presented a fitting function which accurately captures the radial profiles of our simulations (see Eq. \eqref{fitfunc}; Fig. ~\ref{fit}). The free parameter $L_z$ in our fitting function can be thought of as the typical projected radial extent of the \ion{H}{i} halo with respect to the center of haloes. We found that the \ion{H}{i} radial profiles of LLSs in FIREbox are much less extended than those studied in the OWLS and EAGLE simulations \citep{2014MNRAS.438..529R, 2015MNRAS.452.2034R}.
    \item Our FIRE-2 simulations predict cumulative covering fractions for $M_{\mathrm{vir}} = 10^{12} M_{\odot}$ haloes that are in agreement with those observed in $M_{\mathrm{vir}} \approx 10^{12} M_{\odot}$ LBGs at redshifts $z = 2 - 2.5$ by \citet{2012ApJ...750...67R}, see Fig. ~\ref{obs_tot}.
    \item However, our simulations do not reproduce the observations of \citet{2013ApJ...762L..19P} for QSOs hosted in $M_{\mathrm{vir}} \approx 10^{12.5} M_{\odot}$ haloes at redshifts $z= 2 - 2.5$. In particular, the average covering fraction of the observed sample is almost twice as large as the typical value found in our simulations (Fig. ~\ref{obs_tot}).
    \item Comparing the radial plots of \ion{H}{i} covering fraction from FIREbox with the observations indicates that the simulations agree with the distribution inside haloes, but underestimate the extent of \ion{H}{i} in the outer regions (Fig. ~\ref{dflls_obs}). This seems to point at missing \ion{H}{i} at large radii in the simulations. In particular, \citet{2013ApJ...776..136P} find that the \ion{H}{i} radial covering fraction stagnates at around 20\% at all radii, whereas FIRE-2 simulations predict a decay to zero outside the virial radius of $M_{\mathrm{vir}}\sim 10^{12.5} M_{\sun}$ haloes.
    \item We compute the \ion{H}{i} covering fraction of strong absorbers at additional density cuts of $N_{\ion{H}{i}} > 15.5$ cm$^{-2}$, $N_{\ion{H}{i}} > 19.0$ cm$^{-2}$ and $N_{\ion{H}{i}} > 20.3$ cm$^{-2}$, and show that they are in good agreement with observational measurements \citep{2012ApJ...750...67R} and the EAGLE simulations \citep{2021MNRAS.501.4396G} (see Fig. ~\ref{cfnhi}). \fb{}'s high dynamical range allows us to extend the study of such systems to much lower halo masses and in a broader range of redshifts, providing avenues for future comparison with new simulations and observational campaigns.
\end{itemize}

Further work investigating the covering fraction in haloes hosting massive quasars needs to be conducted. This can be realized by examining much larger datasets of very-massive haloes in simulations and exploring the role of different feedback origins. The expanded samples of very-massive FIRE haloes, with and without AGN feedback, introduced in \citet{10.1093/mnras/stad511, 2023arXiv231016086B}, present a substantial opportunity to both extend the study of the \ion{H}{i} covering fraction in FIRE to more massive haloes and to assess the effective contribution of super-massive black holes in shaping the distribution of cool gas in the CGM. Likewise, including feedback from AGNs in future iterations of FIREbox and MassiveFIRE simulations will be an essential complement to the findings presented in this work, albeit at the cost of increased uncertainty in the form of modelling degeneracies.

Measurements of the covering fraction of \ion{H}{i} in lower mass haloes are also needed to more comprehensively compare with the dependence for $M_{\mathrm{vir}} \lesssim 10^{11} M_{\sun}$ haloes presented in our study. Although this remains a challenge, future campaigns promise advances in this regard thanks to considerable strides in instrumental capabilities and analytical methodologies over recent years. For instance, instruments such as MUSE \citep[Multi Unit Spectroscopic Explorer;][]{2010SPIE.7735E..08B} and KCWI \citep[Keck Cosmic Web Imager;][]{2018ApJ...864...93M} are poised to explore absorption lines of gas in the circumgalactic medium and extend the analysis to lower galaxy masses than previously achievable \citep{2020MNRAS.499.5022D}. Additionally, the planned instruments MOSAIC \citep[Multi-Object Spectrograph;][]{2015arXiv150104726E} and ANDES \citep[ArmazoNes high Dispersion Echelle Spectrograph;][]{2022SPIE12184E..24M} on the Extremely Large Telescope \citep{2007Msngr.127...11G, 2018sf2a.conf....3N} are anticipated to offer the highest precision attainable across a great range of objects and redshifts in the coming decades. These advancements open exciting new avenues for comparisons with our research and other studies in the field.

\section*{Acknowledgements}

We thank an anonymous referee for their helpful and constructive report. RF, MB acknowledge financial support from the Swiss National Science Foundation (grant no. 200021\_188552). RF acknowledges financial support from the Swiss National Science Foundation (grant no. PP00P2\_194814). CAFG was supported by NSF through grants AST-2108230, AST-2307327, and CAREER award AST-1652522; by NASA through grant 17-ATP17-0067 and 21-ATP21-0036; by STScI through grant HST-GO-16730.016-A; and by CXO through grant TM2-23005X. We acknowledge PRACE for awarding us access to MareNostrum at the Barcelona Supercomputing Center (BSC), Spain. The FIREbox simulation was supported in part by a computing allocation from the Swiss National Supercomputing Centre (CSCS) under project IDs s697, s698, and uzh18. We acknowledge access to Piz Daint at the Swiss National Supercomputing Centre, Switzerland under the University of Zurich’s share with the project ID uzh18. The authors would like to acknowledge the University of Zurich's Science IT (www.s3it.uzh.ch) team for their support. All plots in this paper were created with the Matplotlib library for visualization with Python \citep{2007CSE.....9...90H}.

\section*{Data Availability}
 
The data underlying this article are available on reasonable request to the corresponding author. A public version of the \textsf{GIZMO} code is available at \url{http://www.tapir.caltech.edu/~phopkins/Site/GIZMO.html}.


\bibliographystyle{mnras}
\bibliography{document}

\begin{thebibliography}{}
\makeatletter
\relax
\def\mn@urlcharsother{\let\do\@makeother \do\$\do\&\do\#\do\^\do\_\do\%\do\~}
\def\mn@doi{\begingroup\mn@urlcharsother \@ifnextchar [ {\mn@doi@}
  {\mn@doi@[]}}
\def\mn@doi@[#1]#2{\def\@tempa{#1}\ifx\@tempa\@empty \href
  {http://dx.doi.org/#2} {doi:#2}\else \href {http://dx.doi.org/#2} {#1}\fi
  \endgroup}
\def\mn@eprint#1#2{\mn@eprint@#1:#2::\@nil}
\def\mn@eprint@arXiv#1{\href {http://arxiv.org/abs/#1} {{\tt arXiv:#1}}}
\def\mn@eprint@dblp#1{\href {http://dblp.uni-trier.de/rec/bibtex/#1.xml}
  {dblp:#1}}
\def\mn@eprint@#1:#2:#3:#4\@nil{\def\@tempa {#1}\def\@tempb {#2}\def\@tempc
  {#3}\ifx \@tempc \@empty \let \@tempc \@tempb \let \@tempb \@tempa \fi \ifx
  \@tempb \@empty \def\@tempb {arXiv}\fi \@ifundefined
  {mn@eprint@\@tempb}{\@tempb:\@tempc}{\expandafter \expandafter \csname
  mn@eprint@\@tempb\endcsname \expandafter{\@tempc}}}

\bibitem[\protect\citeauthoryear{Ade et~al.,}{Ade et~al.}{2016}]{2016}
Ade P. A.~R.,  et~al., 2016, \mn@doi [Astronomy & Astrophysics]
  {10.1051/0004-6361/201525830}, 594, A13

\bibitem[\protect\citeauthoryear{{Altay}, {Theuns}, {Schaye}, {Crighton}  \&
  {Dalla Vecchia}}{{Altay} et~al.}{2011}]{2011ApJ...737L..37A}
{Altay} G.,  {Theuns} T.,  {Schaye} J.,  {Crighton} N. H.~M.,   {Dalla Vecchia}
  C.,  2011, \mn@doi [\apjl] {10.1088/2041-8205/737/2/L37}, \href
  {https://ui.adsabs.harvard.edu/abs/2011ApJ...737L..37A} {737, L37}

\bibitem[\protect\citeauthoryear{{Angl{\'e}s-Alc{\'a}zar},
  {Faucher-Gigu{\`e}re}, {Kere{\v{s}}}, {Hopkins}, {Quataert}  \&
  {Murray}}{{Angl{\'e}s-Alc{\'a}zar} et~al.}{2017}]{2017MNRAS.470.4698A}
{Angl{\'e}s-Alc{\'a}zar} D.,  {Faucher-Gigu{\`e}re} C.-A.,  {Kere{\v{s}}} D.,
  {Hopkins} P.~F.,  {Quataert} E.,   {Murray} N.,  2017, \mn@doi [\mnras]
  {10.1093/mnras/stx1517}, \href
  {https://ui.adsabs.harvard.edu/abs/2017MNRAS.470.4698A} {470, 4698}

\bibitem[\protect\citeauthoryear{{Antonucci}}{{Antonucci}}{1993}]{1993ARA&A..31..473A}
{Antonucci} R.,  1993, \mn@doi [\araa] {10.1146/annurev.aa.31.090193.002353},
  \href {https://ui.adsabs.harvard.edu/abs/1993ARA&A..31..473A} {31, 473}

\bibitem[\protect\citeauthoryear{{Aravena} et~al.,}{{Aravena}
  et~al.}{2016}]{2016ApJ...833...68A}
{Aravena} M.,  et~al., 2016, \mn@doi [\apj] {10.3847/1538-4357/833/1/68}, \href
  {https://ui.adsabs.harvard.edu/abs/2016ApJ...833...68A} {833, 68}

\bibitem[\protect\citeauthoryear{{Ayromlou}, {Nelson}  \&
  {Pillepich}}{{Ayromlou} et~al.}{2023}]{2023MNRAS.524.5391A}
{Ayromlou} M.,  {Nelson} D.,   {Pillepich} A.,  2023, \mn@doi [\mnras]
  {10.1093/mnras/stad2046}, \href
  {https://ui.adsabs.harvard.edu/abs/2023MNRAS.524.5391A} {524, 5391}

\bibitem[\protect\citeauthoryear{{Bacon} et~al.,}{{Bacon}
  et~al.}{2010}]{2010SPIE.7735E..08B}
{Bacon} R.,  et~al., 2010, in {McLean} I.~S.,  {Ramsay} S.~K.,   {Takami} H.,
  eds,  Society of Photo-Optical Instrumentation Engineers (SPIE) Conference
  Series Vol. 7735, Ground-based and Airborne Instrumentation for Astronomy
  III. p. 773508 (\mn@eprint {arXiv} {2211.16795}), \mn@doi{10.1117/12.856027}

\bibitem[\protect\citeauthoryear{{Barnes}, {Kannan}, {Vogelsberger}  \&
  {Marinacci}}{{Barnes} et~al.}{2020}]{2020MNRAS.494.1143B}
{Barnes} D.~J.,  {Kannan} R.,  {Vogelsberger} M.,   {Marinacci} F.,  2020,
  \mn@doi [\mnras] {10.1093/mnras/staa591}, \href
  {https://ui.adsabs.harvard.edu/abs/2020MNRAS.494.1143B} {494, 1143}

\bibitem[\protect\citeauthoryear{{Bauermeister}, {Blitz}  \&
  {Ma}}{{Bauermeister} et~al.}{2010}]{2010ApJ...717..323B}
{Bauermeister} A.,  {Blitz} L.,   {Ma} C.-P.,  2010, \mn@doi [\apj]
  {10.1088/0004-637X/717/1/323}, \href
  {https://ui.adsabs.harvard.edu/abs/2010ApJ...717..323B} {717, 323}

\bibitem[\protect\citeauthoryear{{Biernacki} \& {Teyssier}}{{Biernacki} \&
  {Teyssier}}{2018}]{2018MNRAS.475.5688B}
{Biernacki} P.,  {Teyssier} R.,  2018, \mn@doi [\mnras] {10.1093/mnras/sty216},
  \href {https://ui.adsabs.harvard.edu/abs/2018MNRAS.475.5688B} {475, 5688}

\bibitem[\protect\citeauthoryear{{Binney}}{{Binney}}{1977}]{1977ApJ...215..483B}
{Binney} J.,  1977, \mn@doi [\apj] {10.1086/155378}, \href
  {https://ui.adsabs.harvard.edu/abs/1977ApJ...215..483B} {215, 483}

\bibitem[\protect\citeauthoryear{{Birnboim} \& {Dekel}}{{Birnboim} \&
  {Dekel}}{2003}]{2003MNRAS.345..349B}
{Birnboim} Y.,  {Dekel} A.,  2003, \mn@doi [\mnras]
  {10.1046/j.1365-8711.2003.06955.x}, \href
  {https://ui.adsabs.harvard.edu/abs/2003MNRAS.345..349B} {345, 349}

\bibitem[\protect\citeauthoryear{{Booth}, {de Blok}, {Jonas}  \&
  {Fanaroff}}{{Booth} et~al.}{2009}]{2009arXiv0910.2935B}
{Booth} R.~S.,  {de Blok} W.~J.~G.,  {Jonas} J.~L.,   {Fanaroff} B.,  2009,
  \mn@doi [arXiv e-prints] {10.48550/arXiv.0910.2935}, \href
  {https://ui.adsabs.harvard.edu/abs/2009arXiv0910.2935B} {p. arXiv:0910.2935}

\bibitem[\protect\citeauthoryear{{Bromm}, {Coppi}  \& {Larson}}{{Bromm}
  et~al.}{2002}]{2002ApJ...564...23B}
{Bromm} V.,  {Coppi} P.~S.,   {Larson} R.~B.,  2002, \mn@doi [\apj]
  {10.1086/323947}, \href
  {https://ui.adsabs.harvard.edu/abs/2002ApJ...564...23B} {564, 23}

\bibitem[\protect\citeauthoryear{{Bryan} \& {Norman}}{{Bryan} \&
  {Norman}}{1998}]{1998ApJ...495...80B}
{Bryan} G.~L.,  {Norman} M.~L.,  1998, \mn@doi [\apj] {10.1086/305262}, \href
  {https://ui.adsabs.harvard.edu/abs/1998ApJ...495...80B} {495, 80}

\bibitem[\protect\citeauthoryear{{Byrne} et~al.,}{{Byrne}
  et~al.}{2023}]{2023arXiv231016086B}
{Byrne} L.,  et~al., 2023, \mn@doi [arXiv e-prints]
  {10.48550/arXiv.2310.16086}, \href
  {https://ui.adsabs.harvard.edu/abs/2023arXiv231016086B} {p. arXiv:2310.16086}

\bibitem[\protect\citeauthoryear{{Chen} \& {Mulchaey}}{{Chen} \&
  {Mulchaey}}{2009}]{2009ApJ...701.1219C}
{Chen} H.-W.,  {Mulchaey} J.~S.,  2009, \mn@doi [\apj]
  {10.1088/0004-637X/701/2/1219}, \href
  {https://ui.adsabs.harvard.edu/abs/2009ApJ...701.1219C} {701, 1219}

\bibitem[\protect\citeauthoryear{{Crain} et~al.,}{{Crain}
  et~al.}{2015}]{2015MNRAS.450.1937C}
{Crain} R.~A.,  et~al., 2015, \mn@doi [\mnras] {10.1093/mnras/stv725}, \href
  {https://ui.adsabs.harvard.edu/abs/2015MNRAS.450.1937C} {450, 1937}

\bibitem[\protect\citeauthoryear{Crighton et~al.,}{Crighton
  et~al.}{2015}]{10.1093/mnras/stv1182}
Crighton N. H.~M.,  et~al., 2015, \mn@doi [Monthly Notices of the Royal
  Astronomical Society] {10.1093/mnras/stv1182}, 452, 217

\bibitem[\protect\citeauthoryear{{Dekel} et~al.,}{{Dekel}
  et~al.}{2009}]{2009Natur.457..451D}
{Dekel} A.,  et~al., 2009, \mn@doi [\nat] {10.1038/nature07648}, \href
  {https://ui.adsabs.harvard.edu/abs/2009Natur.457..451D} {457, 451}

\bibitem[\protect\citeauthoryear{{Dessauges-Zavadsky}
  et~al.,}{{Dessauges-Zavadsky} et~al.}{2019}]{2019NatAs...3.1115D}
{Dessauges-Zavadsky} M.,  et~al., 2019, \mn@doi [Nature Astronomy]
  {10.1038/s41550-019-0874-0}, \href
  {https://ui.adsabs.harvard.edu/abs/2019NatAs...3.1115D} {3, 1115}

\bibitem[\protect\citeauthoryear{{Diemer} et~al.,}{{Diemer}
  et~al.}{2019}]{2019MNRAS.487.1529D}
{Diemer} B.,  et~al., 2019, \mn@doi [\mnras] {10.1093/mnras/stz1323}, \href
  {https://ui.adsabs.harvard.edu/abs/2019MNRAS.487.1529D} {487, 1529}

\bibitem[\protect\citeauthoryear{Dutta}{Dutta}{2019}]{Dutta_2019}
Dutta R.,  2019, \mn@doi [Journal of Astrophysics and Astronomy]
  {10.1007/s12036-019-9610-5}, 40

\bibitem[\protect\citeauthoryear{{Dutta} et~al.,}{{Dutta}
  et~al.}{2020}]{2020MNRAS.499.5022D}
{Dutta} R.,  et~al., 2020, \mn@doi [\mnras] {10.1093/mnras/staa3147}, \href
  {https://ui.adsabs.harvard.edu/abs/2020MNRAS.499.5022D} {499, 5022}

\bibitem[\protect\citeauthoryear{{Efstathiou} \& {Silk}}{{Efstathiou} \&
  {Silk}}{1983}]{1983FCPh....9....1E}
{Efstathiou} G.,  {Silk} J.,  1983, \fcp, \href
  {https://ui.adsabs.harvard.edu/abs/1983FCPh....9....1E} {9, 1}

\bibitem[\protect\citeauthoryear{{Evans} et~al.,}{{Evans}
  et~al.}{2015}]{2015arXiv150104726E}
{Evans} C.,  et~al., 2015, \mn@doi [arXiv e-prints]
  {10.48550/arXiv.1501.04726}, \href
  {https://ui.adsabs.harvard.edu/abs/2015arXiv150104726E} {p. arXiv:1501.04726}

\bibitem[\protect\citeauthoryear{{Faucher-Gigu{\`e}re} \&
  {Kere{\v{s}}}}{{Faucher-Gigu{\`e}re} \&
  {Kere{\v{s}}}}{2011}]{2011MNRAS.412L.118F}
{Faucher-Gigu{\`e}re} C.-A.,  {Kere{\v{s}}} D.,  2011, \mn@doi [\mnras]
  {10.1111/j.1745-3933.2011.01018.x}, \href
  {https://ui.adsabs.harvard.edu/abs/2011MNRAS.412L.118F} {412, L118}

\bibitem[\protect\citeauthoryear{{Faucher-Gigu{\`e}re} \&
  {Oh}}{{Faucher-Gigu{\`e}re} \& {Oh}}{2023}]{2023ARA&A..61..131F}
{Faucher-Gigu{\`e}re} C.-A.,  {Oh} S.~P.,  2023, \mn@doi [\araa]
  {10.1146/annurev-astro-052920-125203}, \href
  {https://ui.adsabs.harvard.edu/abs/2023ARA&A..61..131F} {61, 131}

\bibitem[\protect\citeauthoryear{{Faucher-Gigu{\`e}re}, {Kere{\v{s}}},
  {Dijkstra}, {Hernquist}  \& {Zaldarriaga}}{{Faucher-Gigu{\`e}re}
  et~al.}{2010}]{2010ApJ...725..633F}
{Faucher-Gigu{\`e}re} C.-A.,  {Kere{\v{s}}} D.,  {Dijkstra} M.,  {Hernquist}
  L.,   {Zaldarriaga} M.,  2010, \mn@doi [\apj] {10.1088/0004-637X/725/1/633},
  \href {https://ui.adsabs.harvard.edu/abs/2010ApJ...725..633F} {725, 633}

\bibitem[\protect\citeauthoryear{{Faucher-Gigu{\`e}re}, {Kere{\v{s}}}  \&
  {Ma}}{{Faucher-Gigu{\`e}re} et~al.}{2011}]{2011MNRAS.417.2982F}
{Faucher-Gigu{\`e}re} C.-A.,  {Kere{\v{s}}} D.,   {Ma} C.-P.,  2011, \mn@doi
  [\mnras] {10.1111/j.1365-2966.2011.19457.x}, \href
  {https://ui.adsabs.harvard.edu/abs/2011MNRAS.417.2982F} {417, 2982}

\bibitem[\protect\citeauthoryear{{Faucher-Gigu{\`e}re}, {Hopkins},
  {Kere{\v{s}}}, {Muratov}, {Quataert}  \& {Murray}}{{Faucher-Gigu{\`e}re}
  et~al.}{2015}]{2015MNRAS.449..987F}
{Faucher-Gigu{\`e}re} C.-A.,  {Hopkins} P.~F.,  {Kere{\v{s}}} D.,  {Muratov}
  A.~L.,  {Quataert} E.,   {Murray} N.,  2015, \mn@doi [\mnras]
  {10.1093/mnras/stv336}, \href
  {https://ui.adsabs.harvard.edu/abs/2015MNRAS.449..987F} {449, 987}

\bibitem[\protect\citeauthoryear{{Faucher-Gigu{\`e}re}, {Feldmann}, {Quataert},
  {Kere{\v{s}}}, {Hopkins}  \& {Murray}}{{Faucher-Gigu{\`e}re}
  et~al.}{2016}]{2016MNRAS.461L..32F}
{Faucher-Gigu{\`e}re} C.-A.,  {Feldmann} R.,  {Quataert} E.,  {Kere{\v{s}}} D.,
   {Hopkins} P.~F.,   {Murray} N.,  2016, \mn@doi [\mnras]
  {10.1093/mnrasl/slw091}, \href
  {https://ui.adsabs.harvard.edu/abs/2016MNRAS.461L..32F} {461, L32}

\bibitem[\protect\citeauthoryear{{Feldmann}}{{Feldmann}}{2020}]{2020CmPhy...3..226F}
{Feldmann} R.,  2020, \mn@doi [Communications Physics]
  {10.1038/s42005-020-00493-0}, \href
  {https://ui.adsabs.harvard.edu/abs/2020CmPhy...3..226F} {3, 226}

\bibitem[\protect\citeauthoryear{{Feldmann}, {Hopkins}, {Quataert},
  {Faucher-Gigu{\`e}re}  \& {Kere{\v{s}}}}{{Feldmann}
  et~al.}{2016}]{2016MNRAS.458L..14F}
{Feldmann} R.,  {Hopkins} P.~F.,  {Quataert} E.,  {Faucher-Gigu{\`e}re} C.-A.,
   {Kere{\v{s}}} D.,  2016, \mn@doi [\mnras] {10.1093/mnrasl/slw014}, \href
  {https://ui.adsabs.harvard.edu/abs/2016MNRAS.458L..14F} {458, L14}

\bibitem[\protect\citeauthoryear{{Feldmann}, {Quataert}, {Hopkins},
  {Faucher-Gigu{\`e}re}  \& {Kere{\v{s}}}}{{Feldmann}
  et~al.}{2017}]{2017MNRAS.470.1050F}
{Feldmann} R.,  {Quataert} E.,  {Hopkins} P.~F.,  {Faucher-Gigu{\`e}re} C.-A.,
   {Kere{\v{s}}} D.,  2017, \mn@doi [\mnras] {10.1093/mnras/stx1120}, \href
  {https://ui.adsabs.harvard.edu/abs/2017MNRAS.470.1050F} {470, 1050}

\bibitem[\protect\citeauthoryear{{Feldmann} et~al.,}{{Feldmann}
  et~al.}{2023}]{2023MNRAS.522.3831F}
{Feldmann} R.,  et~al., 2023, \mn@doi [\mnras] {10.1093/mnras/stad1205}, \href
  {https://ui.adsabs.harvard.edu/abs/2023MNRAS.522.3831F} {522, 3831}

\bibitem[\protect\citeauthoryear{{Ferland}, {Korista}, {Verner}, {Ferguson},
  {Kingdon}  \& {Verner}}{{Ferland} et~al.}{1998}]{1998PASP..110..761F}
{Ferland} G.~J.,  {Korista} K.~T.,  {Verner} D.~A.,  {Ferguson} J.~W.,
  {Kingdon} J.~B.,   {Verner} E.~M.,  1998, \mn@doi [\pasp] {10.1086/316190},
  \href {https://ui.adsabs.harvard.edu/abs/1998PASP..110..761F} {110, 761}

\bibitem[\protect\citeauthoryear{{Findlay} et~al.,}{{Findlay}
  et~al.}{2018}]{2018ApJS..236...44F}
{Findlay} J.~R.,  et~al., 2018, \mn@doi [\apjs] {10.3847/1538-4365/aabee5},
  \href {https://ui.adsabs.harvard.edu/abs/2018ApJS..236...44F} {236, 44}

\bibitem[\protect\citeauthoryear{{Fumagalli}, {Prochaska}, {Kasen}, {Dekel},
  {Ceverino}  \& {Primack}}{{Fumagalli} et~al.}{2011}]{2011MNRAS.418.1796F}
{Fumagalli} M.,  {Prochaska} J.~X.,  {Kasen} D.,  {Dekel} A.,  {Ceverino} D.,
  {Primack} J.~R.,  2011, \mn@doi [\mnras] {10.1111/j.1365-2966.2011.19599.x},
  \href {https://ui.adsabs.harvard.edu/abs/2011MNRAS.418.1796F} {418, 1796}

\bibitem[\protect\citeauthoryear{{Fumagalli}, {Hennawi}, {Prochaska}, {Kasen},
  {Dekel}, {Ceverino}  \& {Primack}}{{Fumagalli}
  et~al.}{2014}]{2014ApJ...780...74F}
{Fumagalli} M.,  {Hennawi} J.~F.,  {Prochaska} J.~X.,  {Kasen} D.,  {Dekel} A.,
   {Ceverino} D.,   {Primack} J.,  2014, \mn@doi [\apj]
  {10.1088/0004-637X/780/1/74}, \href
  {https://ui.adsabs.harvard.edu/abs/2014ApJ...780...74F} {780, 74}

\bibitem[\protect\citeauthoryear{{Garratt-Smithson}, {Power}, {Lagos},
  {Stevens}, {Allison}  \& {Sadler}}{{Garratt-Smithson}
  et~al.}{2021}]{2021MNRAS.501.4396G}
{Garratt-Smithson} L.,  {Power} C.,  {Lagos} C. d.~P.,  {Stevens} A. R.~H.,
  {Allison} J.~R.,   {Sadler} E.~M.,  2021, \mn@doi [\mnras]
  {10.1093/mnras/staa3870}, \href
  {https://ui.adsabs.harvard.edu/abs/2021MNRAS.501.4396G} {501, 4396}

\bibitem[\protect\citeauthoryear{{Gensior}, {Feldmann}, {Mayer}, {Wetzel},
  {Hopkins}  \& {Faucher-Gigu{\`e}re}}{{Gensior} et~al.}{2023}]{gensior+2023}
{Gensior} J.,  {Feldmann} R.,  {Mayer} L.,  {Wetzel} A.,  {Hopkins} P.~F.,
  {Faucher-Gigu{\`e}re} C.-A.,  2023, \mn@doi [\mnras]
  {10.1093/mnrasl/slac138}, \href
  {https://ui.adsabs.harvard.edu/abs/2023MNRAS.518L..63G} {518, L63}

\bibitem[\protect\citeauthoryear{{Gill}, {Knebe}  \& {Gibson}}{{Gill}
  et~al.}{2004}]{2004MNRAS.351..399G}
{Gill} S. P.~D.,  {Knebe} A.,   {Gibson} B.~K.,  2004, \mn@doi [\mnras]
  {10.1111/j.1365-2966.2004.07786.x}, \href
  {https://ui.adsabs.harvard.edu/abs/2004MNRAS.351..399G} {351, 399}

\bibitem[\protect\citeauthoryear{{Gilmozzi} \& {Spyromilio}}{{Gilmozzi} \&
  {Spyromilio}}{2007}]{2007Msngr.127...11G}
{Gilmozzi} R.,  {Spyromilio} J.,  2007, The Messenger, \href
  {https://ui.adsabs.harvard.edu/abs/2007Msngr.127...11G} {127, 11}

\bibitem[\protect\citeauthoryear{{Glowacki} et~al.,}{{Glowacki}
  et~al.}{2019}]{2019MNRAS.489.4926G}
{Glowacki} M.,  et~al., 2019, \mn@doi [\mnras] {10.1093/mnras/stz2452}, \href
  {https://ui.adsabs.harvard.edu/abs/2019MNRAS.489.4926G} {489, 4926}

\bibitem[\protect\citeauthoryear{{Guglielmo}, {Poggianti}, {Moretti}, {Fritz},
  {Calvi}, {Vulcani}, {Fasano}  \& {Paccagnella}}{{Guglielmo}
  et~al.}{2015}]{2015MNRAS.450.2749G}
{Guglielmo} V.,  {Poggianti} B.~M.,  {Moretti} A.,  {Fritz} J.,  {Calvi} R.,
  {Vulcani} B.,  {Fasano} G.,   {Paccagnella} A.,  2015, \mn@doi [\mnras]
  {10.1093/mnras/stv757}, \href
  {https://ui.adsabs.harvard.edu/abs/2015MNRAS.450.2749G} {450, 2749}

\bibitem[\protect\citeauthoryear{{Guo}, {White}, {Li}  \&
  {Boylan-Kolchin}}{{Guo} et~al.}{2010}]{2010MNRAS.404.1111G}
{Guo} Q.,  {White} S.,  {Li} C.,   {Boylan-Kolchin} M.,  2010, \mn@doi [\mnras]
  {10.1111/j.1365-2966.2010.16341.x}, \href
  {https://ui.adsabs.harvard.edu/abs/2010MNRAS.404.1111G} {404, 1111}

\bibitem[\protect\citeauthoryear{{Gutcke}, {Stinson}, {Macci{\`o}}, {Wang}  \&
  {Dutton}}{{Gutcke} et~al.}{2017}]{2017MNRAS.464.2796G}
{Gutcke} T.~A.,  {Stinson} G.~S.,  {Macci{\`o}} A.~V.,  {Wang} L.,   {Dutton}
  A.~A.,  2017, \mn@doi [\mnras] {10.1093/mnras/stw2539}, \href
  {https://ui.adsabs.harvard.edu/abs/2017MNRAS.464.2796G} {464, 2796}

\bibitem[\protect\citeauthoryear{Hahn \& Abel}{Hahn \& Abel}{2011}]{2011}
Hahn O.,  Abel T.,  2011, \mn@doi [Monthly Notices of the Royal Astronomical
  Society] {10.1111/j.1365-2966.2011.18820.x}, 415, 2101–2121

\bibitem[\protect\citeauthoryear{{Hayashi} \& {Nakano}}{{Hayashi} \&
  {Nakano}}{1965}]{1965PThPh..34..754H}
{Hayashi} C.,  {Nakano} T.,  1965, \mn@doi [Progress of Theoretical Physics]
  {10.1143/PTP.34.754}, \href
  {https://ui.adsabs.harvard.edu/abs/1965PThPh..34..754H} {34, 754}

\bibitem[\protect\citeauthoryear{{Ho}, {Martin}  \& {Turner}}{{Ho}
  et~al.}{2019}]{2019ApJ...875...54H}
{Ho} S.~H.,  {Martin} C.~L.,   {Turner} M.~L.,  2019, \mn@doi [\apj]
  {10.3847/1538-4357/ab0ec2}, \href
  {https://ui.adsabs.harvard.edu/abs/2019ApJ...875...54H} {875, 54}

\bibitem[\protect\citeauthoryear{Hopkins}{Hopkins}{2015}]{10.1093/mnras/stv195}
Hopkins P.~F.,  2015, \mn@doi [Monthly Notices of the Royal Astronomical
  Society] {10.1093/mnras/stv195}, 450, 53

\bibitem[\protect\citeauthoryear{{Hopkins} \& {Grudi{\'c}}}{{Hopkins} \&
  {Grudi{\'c}}}{2019}]{2019MNRAS.483.4187H}
{Hopkins} P.~F.,  {Grudi{\'c}} M.~Y.,  2019, \mn@doi [\mnras]
  {10.1093/mnras/sty3089}, \href
  {https://ui.adsabs.harvard.edu/abs/2019MNRAS.483.4187H} {483, 4187}

\bibitem[\protect\citeauthoryear{{Hopkins}, {Quataert}  \& {Murray}}{{Hopkins}
  et~al.}{2012}]{2012MNRAS.421.3488H}
{Hopkins} P.~F.,  {Quataert} E.,   {Murray} N.,  2012, \mn@doi [\mnras]
  {10.1111/j.1365-2966.2012.20578.x}, \href
  {https://ui.adsabs.harvard.edu/abs/2012MNRAS.421.3488H} {421, 3488}

\bibitem[\protect\citeauthoryear{{Hopkins}, Kereš, Oñorbe, Faucher-Giguère,
  Quataert, Murray  \& Bullock}{{Hopkins} et~al.}{2014}]{2014}
{Hopkins} P.~F.,  Kereš D.,  Oñorbe J.,  Faucher-Giguère C.-A.,  Quataert
  E.,  Murray N.,   Bullock J.~S.,  2014, \mn@doi [Monthly Notices of the Royal
  Astronomical Society] {10.1093/mnras/stu1738}, 445, 581–603

\bibitem[\protect\citeauthoryear{{Hopkins} et~al.,}{{Hopkins}
  et~al.}{2018}]{2018MNRAS.480..800H}
{Hopkins} P.~F.,  et~al., 2018, \mn@doi [\mnras] {10.1093/mnras/sty1690}, \href
  {https://ui.adsabs.harvard.edu/abs/2018MNRAS.480..800H} {480, 800}

\bibitem[\protect\citeauthoryear{{Hopkins} et~al.,}{{Hopkins}
  et~al.}{2023}]{2023MNRAS.519.3154H}
{Hopkins} P.~F.,  et~al., 2023, \mn@doi [\mnras] {10.1093/mnras/stac3489},
  \href {https://ui.adsabs.harvard.edu/abs/2023MNRAS.519.3154H} {519, 3154}

\bibitem[\protect\citeauthoryear{{Hunter}}{{Hunter}}{2007}]{2007CSE.....9...90H}
{Hunter} J.~D.,  2007, \mn@doi [Computing in Science and Engineering]
  {10.1109/MCSE.2007.55}, \href
  {https://ui.adsabs.harvard.edu/abs/2007CSE.....9...90H} {9, 90}

\bibitem[\protect\citeauthoryear{{Jaiswal} \& {Omar}}{{Jaiswal} \&
  {Omar}}{2020}]{2020MNRAS.498.4745J}
{Jaiswal} S.,  {Omar} A.,  2020, \mn@doi [\mnras] {10.1093/mnras/staa2420},
  \href {https://ui.adsabs.harvard.edu/abs/2020MNRAS.498.4745J} {498, 4745}

\bibitem[\protect\citeauthoryear{{Jonas} \& {MeerKAT Team}}{{Jonas} \& {MeerKAT
  Team}}{2016}]{2016mks..confE...1J}
{Jonas} J.,  {MeerKAT Team} 2016, in MeerKAT Science: On the Pathway to the
  SKA. p.~1, \mn@doi{10.22323/1.277.0001}

\bibitem[\protect\citeauthoryear{{Kere{\v{s}}}, {Katz}, {Weinberg}  \&
  {Dav{\'e}}}{{Kere{\v{s}}} et~al.}{2005}]{2005MNRAS.363....2K}
{Kere{\v{s}}} D.,  {Katz} N.,  {Weinberg} D.~H.,   {Dav{\'e}} R.,  2005,
  \mn@doi [\mnras] {10.1111/j.1365-2966.2005.09451.x}, \href
  {https://ui.adsabs.harvard.edu/abs/2005MNRAS.363....2K} {363, 2}

\bibitem[\protect\citeauthoryear{{Khrykin}, {Sorini}, {Lee}  \&
  {Dav{\'e}}}{{Khrykin} et~al.}{2024}]{2024MNRAS.529..537K}
{Khrykin} I.~S.,  {Sorini} D.,  {Lee} K.-G.,   {Dav{\'e}} R.,  2024, \mn@doi
  [\mnras] {10.1093/mnras/stae525}, \href
  {https://ui.adsabs.harvard.edu/abs/2024MNRAS.529..537K} {529, 537}

\bibitem[\protect\citeauthoryear{{Kirby}, {Koribalski}, {Jerjen}  \&
  {L{\'o}pez-S{\'a}nchez}}{{Kirby} et~al.}{2012}]{2012MNRAS.420.2924K}
{Kirby} E.~M.,  {Koribalski} B.,  {Jerjen} H.,   {L{\'o}pez-S{\'a}nchez}
  {\'A}.,  2012, \mn@doi [\mnras] {10.1111/j.1365-2966.2011.20103.x}, \href
  {https://ui.adsabs.harvard.edu/abs/2012MNRAS.420.2924K} {420, 2924}

\bibitem[\protect\citeauthoryear{{Knollmann} \& {Knebe}}{{Knollmann} \&
  {Knebe}}{2009}]{2009ApJS..182..608K}
{Knollmann} S.~R.,  {Knebe} A.,  2009, \mn@doi [\apjs]
  {10.1088/0067-0049/182/2/608}, \href
  {https://ui.adsabs.harvard.edu/abs/2009ApJS..182..608K} {182, 608}

\bibitem[\protect\citeauthoryear{{Le F{\`e}vre} et~al.,}{{Le F{\`e}vre}
  et~al.}{2020}]{2020A&A...643A...1L}
{Le F{\`e}vre} O.,  et~al., 2020, \mn@doi [\aap] {10.1051/0004-6361/201936965},
  \href {https://ui.adsabs.harvard.edu/abs/2020A&A...643A...1L} {643, A1}

\bibitem[\protect\citeauthoryear{{Marconi} et~al.,}{{Marconi}
  et~al.}{2022}]{2022SPIE12184E..24M}
{Marconi} A.,  et~al., 2022, in {Evans} C.~J.,  {Bryant} J.~J.,   {Motohara}
  K.,  eds,  Society of Photo-Optical Instrumentation Engineers (SPIE)
  Conference Series Vol. 12184, Ground-based and Airborne Instrumentation for
  Astronomy IX. p. 1218424, \mn@doi{10.1117/12.2628689}

\bibitem[\protect\citeauthoryear{{Meiksin}, {Bolton}  \& {Tittley}}{{Meiksin}
  et~al.}{2015}]{2015MNRAS.453..899M}
{Meiksin} A.,  {Bolton} J.~S.,   {Tittley} E.~R.,  2015, \mn@doi [\mnras]
  {10.1093/mnras/stv1682}, \href
  {https://ui.adsabs.harvard.edu/abs/2015MNRAS.453..899M} {453, 899}

\bibitem[\protect\citeauthoryear{{Meiksin}, {Bolton}  \& {Puchwein}}{{Meiksin}
  et~al.}{2017}]{2017MNRAS.468.1893M}
{Meiksin} A.,  {Bolton} J.~S.,   {Puchwein} E.,  2017, \mn@doi [\mnras]
  {10.1093/mnras/stx191}, \href
  {https://ui.adsabs.harvard.edu/abs/2017MNRAS.468.1893M} {468, 1893}

\bibitem[\protect\citeauthoryear{{Morrissey} et~al.,}{{Morrissey}
  et~al.}{2018}]{2018ApJ...864...93M}
{Morrissey} P.,  et~al., 2018, \mn@doi [\apj] {10.3847/1538-4357/aad597}, \href
  {https://ui.adsabs.harvard.edu/abs/2018ApJ...864...93M} {864, 93}

\bibitem[\protect\citeauthoryear{{Neichel}, {Mouillet}, {Gendron}, {Correia},
  {Sauvage}  \& {Fusco}}{{Neichel} et~al.}{2018}]{2018sf2a.conf....3N}
{Neichel} B.,  {Mouillet} D.,  {Gendron} E.,  {Correia} C.,  {Sauvage} J.~F.,
  {Fusco} T.,  2018, in {Di Matteo} P.,  {Billebaud} F.,  {Herpin} F.,
  {Lagarde} N.,  {Marquette} J.~B.,  {Robin} A.,   {Venot} O.,  eds, SF2A-2018:
  Proceedings of the Annual meeting of the French Society of Astronomy and
  Astrophysics. p.~Di (\mn@eprint {arXiv} {1812.06639}),
  \mn@doi{10.48550/arXiv.1812.06639}

\bibitem[\protect\citeauthoryear{{Nelson}, {Vogelsberger}, {Genel}, {Sijacki},
  {Kere{\v{s}}}, {Springel}  \& {Hernquist}}{{Nelson}
  et~al.}{2013}]{2013MNRAS.429.3353N}
{Nelson} D.,  {Vogelsberger} M.,  {Genel} S.,  {Sijacki} D.,  {Kere{\v{s}}} D.,
   {Springel} V.,   {Hernquist} L.,  2013, \mn@doi [\mnras]
  {10.1093/mnras/sts595}, \href
  {https://ui.adsabs.harvard.edu/abs/2013MNRAS.429.3353N} {429, 3353}

\bibitem[\protect\citeauthoryear{{Nelson} et~al.,}{{Nelson}
  et~al.}{2020}]{2020MNRAS.498.2391N}
{Nelson} D.,  et~al., 2020, \mn@doi [\mnras] {10.1093/mnras/staa2419}, \href
  {https://ui.adsabs.harvard.edu/abs/2020MNRAS.498.2391N} {498, 2391}

\bibitem[\protect\citeauthoryear{{Noterdaeme} et~al.,}{{Noterdaeme}
  et~al.}{2012}]{2012A&A...547L...1N}
{Noterdaeme} P.,  et~al., 2012, \mn@doi [\aap] {10.1051/0004-6361/201220259},
  \href {https://ui.adsabs.harvard.edu/abs/2012A&A...547L...1N} {547, L1}

\bibitem[\protect\citeauthoryear{Padmanabhan}{Padmanabhan}{2017}]{padmanabhan_2017}
Padmanabhan H.,  2017, \mn@doi [Proceedings of the International Astronomical
  Union] {10.1017/S1743921317010821}, 12, 216–221

\bibitem[\protect\citeauthoryear{{Padovani} et~al.,}{{Padovani}
  et~al.}{2017}]{2017A&ARv..25....2P}
{Padovani} P.,  et~al., 2017, \mn@doi [\aapr] {10.1007/s00159-017-0102-9},
  \href {https://ui.adsabs.harvard.edu/abs/2017A&ARv..25....2P} {25, 2}

\bibitem[\protect\citeauthoryear{{Peebles}}{{Peebles}}{1980}]{1980lssu.book.....P}
{Peebles} P.~J.~E.,  1980, {The large-scale structure of the universe}

\bibitem[\protect\citeauthoryear{{Prochaska}, {Hennawi}  \&
  {Simcoe}}{{Prochaska} et~al.}{2013a}]{2013ApJ...762L..19P}
{Prochaska} J.~X.,  {Hennawi} J.~F.,   {Simcoe} R.~A.,  2013a, \mn@doi [\apjl]
  {10.1088/2041-8205/762/2/L19}, \href
  {https://ui.adsabs.harvard.edu/abs/2013ApJ...762L..19P} {762, L19}

\bibitem[\protect\citeauthoryear{{Prochaska} et~al.,}{{Prochaska}
  et~al.}{2013b}]{2013ApJ...776..136P}
{Prochaska} J.~X.,  et~al., 2013b, \mn@doi [\apj]
  {10.1088/0004-637X/776/2/136}, \href
  {https://ui.adsabs.harvard.edu/abs/2013ApJ...776..136P} {776, 136}

\bibitem[\protect\citeauthoryear{{Prochaska} et~al.,}{{Prochaska}
  et~al.}{2017}]{2017ApJ...837..169P}
{Prochaska} J.~X.,  et~al., 2017, \mn@doi [\apj] {10.3847/1538-4357/aa6007},
  \href {https://ui.adsabs.harvard.edu/abs/2017ApJ...837..169P} {837, 169}

\bibitem[\protect\citeauthoryear{{Rahmati} \& {Schaye}}{{Rahmati} \&
  {Schaye}}{2014}]{2014MNRAS.438..529R}
{Rahmati} A.,  {Schaye} J.,  2014, \mn@doi [\mnras] {10.1093/mnras/stt2235},
  \href {https://ui.adsabs.harvard.edu/abs/2014MNRAS.438..529R} {438, 529}

\bibitem[\protect\citeauthoryear{{Rahmati}, {Schaye}, {Pawlik}  \&
  {Rai{\v{c}}evi{\'c}}}{{Rahmati} et~al.}{2013}]{2013MNRAS.431.2261R}
{Rahmati} A.,  {Schaye} J.,  {Pawlik} A.~H.,   {Rai{\v{c}}evi{\'c}} M.,  2013,
  \mn@doi [\mnras] {10.1093/mnras/stt324}, \href
  {https://ui.adsabs.harvard.edu/abs/2013MNRAS.431.2261R} {431, 2261}

\bibitem[\protect\citeauthoryear{{Rahmati}, {Schaye}, {Bower}, {Crain},
  {Furlong}, {Schaller}  \& {Theuns}}{{Rahmati}
  et~al.}{2015}]{2015MNRAS.452.2034R}
{Rahmati} A.,  {Schaye} J.,  {Bower} R.~G.,  {Crain} R.~A.,  {Furlong} M.,
  {Schaller} M.,   {Theuns} T.,  2015, \mn@doi [\mnras]
  {10.1093/mnras/stv1414}, \href
  {https://ui.adsabs.harvard.edu/abs/2015MNRAS.452.2034R} {452, 2034}

\bibitem[\protect\citeauthoryear{{Rakic}, {Schaye}, {Steidel}  \&
  {Rudie}}{{Rakic} et~al.}{2012}]{2012ApJ...751...94R}
{Rakic} O.,  {Schaye} J.,  {Steidel} C.~C.,   {Rudie} G.~C.,  2012, \mn@doi
  [\apj] {10.1088/0004-637X/751/2/94}, \href
  {https://ui.adsabs.harvard.edu/abs/2012ApJ...751...94R} {751, 94}

\bibitem[\protect\citeauthoryear{Ramesh \& Nelson}{Ramesh \&
  Nelson}{2023}]{ramesh2023zooming}
Ramesh R.,  Nelson D.,  2023, Zooming in on the circumgalactic medium:
  resolving small-scale gas structure with the GIBLE cosmological simulations
  (\mn@eprint {arXiv} {2307.11143})

\bibitem[\protect\citeauthoryear{{Rees} \& {Ostriker}}{{Rees} \&
  {Ostriker}}{1977}]{1977MNRAS.179..541R}
{Rees} M.~J.,  {Ostriker} J.~P.,  1977, \mn@doi [\mnras]
  {10.1093/mnras/179.4.541}, \href
  {https://ui.adsabs.harvard.edu/abs/1977MNRAS.179..541R} {179, 541}

\bibitem[\protect\citeauthoryear{{Reeves}, {Sadler}, {Allison}, {Koribalski},
  {Curran}  \& {Pracy}}{{Reeves} et~al.}{2015}]{2015MNRAS.450..926R}
{Reeves} S.~N.,  {Sadler} E.~M.,  {Allison} J.~R.,  {Koribalski} B.~S.,
  {Curran} S.~J.,   {Pracy} M.~B.,  2015, \mn@doi [\mnras]
  {10.1093/mnras/stv504}, \href
  {https://ui.adsabs.harvard.edu/abs/2015MNRAS.450..926R} {450, 926}

\bibitem[\protect\citeauthoryear{{Rubin}, {Hennawi}, {Prochaska}, {Simcoe},
  {Myers}  \& {Lau}}{{Rubin} et~al.}{2015}]{2015ApJ...808...38R}
{Rubin} K. H.~R.,  {Hennawi} J.~F.,  {Prochaska} J.~X.,  {Simcoe} R.~A.,
  {Myers} A.,   {Lau} M.~W.,  2015, \mn@doi [\apj]
  {10.1088/0004-637X/808/1/38}, \href
  {https://ui.adsabs.harvard.edu/abs/2015ApJ...808...38R} {808, 38}

\bibitem[\protect\citeauthoryear{{Rudie} et~al.,}{{Rudie}
  et~al.}{2012}]{2012ApJ...750...67R}
{Rudie} G.~C.,  et~al., 2012, \mn@doi [\apj] {10.1088/0004-637X/750/1/67},
  \href {https://ui.adsabs.harvard.edu/abs/2012ApJ...750...67R} {750, 67}

\bibitem[\protect\citeauthoryear{{Schaye} et~al.,}{{Schaye}
  et~al.}{2015}]{2015MNRAS.446..521S}
{Schaye} J.,  et~al., 2015, \mn@doi [\mnras] {10.1093/mnras/stu2058}, \href
  {https://ui.adsabs.harvard.edu/abs/2015MNRAS.446..521S} {446, 521}

\bibitem[\protect\citeauthoryear{{Shen}, {Madau}, {Guedes}, {Mayer},
  {Prochaska}  \& {Wadsley}}{{Shen} et~al.}{2013}]{2013ApJ...765...89S}
{Shen} S.,  {Madau} P.,  {Guedes} J.,  {Mayer} L.,  {Prochaska} J.~X.,
  {Wadsley} J.,  2013, \mn@doi [\apj] {10.1088/0004-637X/765/2/89}, \href
  {https://ui.adsabs.harvard.edu/abs/2013ApJ...765...89S} {765, 89}

\bibitem[\protect\citeauthoryear{{Sorini}, {Dav{\'e}}  \&
  {Angl{\'e}s-Alc{\'a}zar}}{{Sorini} et~al.}{2020}]{2020MNRAS.499.2760S}
{Sorini} D.,  {Dav{\'e}} R.,   {Angl{\'e}s-Alc{\'a}zar} D.,  2020, \mn@doi
  [\mnras] {10.1093/mnras/staa2937}, \href
  {https://ui.adsabs.harvard.edu/abs/2020MNRAS.499.2760S} {499, 2760}

\bibitem[\protect\citeauthoryear{{Steidel} et~al.,}{{Steidel}
  et~al.}{2014}]{2014ApJ...795..165S}
{Steidel} C.~C.,  et~al., 2014, \mn@doi [\apj] {10.1088/0004-637X/795/2/165},
  \href {https://ui.adsabs.harvard.edu/abs/2014ApJ...795..165S} {795, 165}

\bibitem[\protect\citeauthoryear{{Stern}, {Fielding}, {Faucher-Gigu{\`e}re}  \&
  {Quataert}}{{Stern} et~al.}{2020}]{2020MNRAS.492.6042S}
{Stern} J.,  {Fielding} D.,  {Faucher-Gigu{\`e}re} C.-A.,   {Quataert} E.,
  2020, \mn@doi [\mnras] {10.1093/mnras/staa198}, \href
  {https://ui.adsabs.harvard.edu/abs/2020MNRAS.492.6042S} {492, 6042}

\bibitem[\protect\citeauthoryear{{Stern} et~al.,}{{Stern}
  et~al.}{2021}]{2021MNRAS.507.2869S}
{Stern} J.,  et~al., 2021, \mn@doi [\mnras] {10.1093/mnras/stab2240}, \href
  {https://ui.adsabs.harvard.edu/abs/2021MNRAS.507.2869S} {507, 2869}

\bibitem[\protect\citeauthoryear{{Strom}, {Steidel}, {Rudie}, {Trainor},
  {Pettini}  \& {Reddy}}{{Strom} et~al.}{2017}]{2017ApJ...836..164S}
{Strom} A.~L.,  {Steidel} C.~C.,  {Rudie} G.~C.,  {Trainor} R.~F.,  {Pettini}
  M.,   {Reddy} N.~A.,  2017, \mn@doi [\apj] {10.3847/1538-4357/836/2/164},
  \href {https://ui.adsabs.harvard.edu/abs/2017ApJ...836..164S} {836, 164}

\bibitem[\protect\citeauthoryear{{Suresh}, {Bird}, {Vogelsberger}, {Genel},
  {Torrey}, {Sijacki}, {Springel}  \& {Hernquist}}{{Suresh}
  et~al.}{2015}]{2015MNRAS.448..895S}
{Suresh} J.,  {Bird} S.,  {Vogelsberger} M.,  {Genel} S.,  {Torrey} P.,
  {Sijacki} D.,  {Springel} V.,   {Hernquist} L.,  2015, \mn@doi [\mnras]
  {10.1093/mnras/stu2762}, \href
  {https://ui.adsabs.harvard.edu/abs/2015MNRAS.448..895S} {448, 895}

\bibitem[\protect\citeauthoryear{{Suresh}, {Nelson}, {Genel}, {Rubin}  \&
  {Hernquist}}{{Suresh} et~al.}{2019}]{2019MNRAS.483.4040S}
{Suresh} J.,  {Nelson} D.,  {Genel} S.,  {Rubin} K. H.~R.,   {Hernquist} L.,
  2019, \mn@doi [\mnras] {10.1093/mnras/sty3402}, \href
  {https://ui.adsabs.harvard.edu/abs/2019MNRAS.483.4040S} {483, 4040}

\bibitem[\protect\citeauthoryear{{Tacconi}, {Genzel}  \& {Sternberg}}{{Tacconi}
  et~al.}{2020}]{2020ARA&A..58..157T}
{Tacconi} L.~J.,  {Genzel} R.,   {Sternberg} A.,  2020, \mn@doi [\araa]
  {10.1146/annurev-astro-082812-141034}, \href
  {https://ui.adsabs.harvard.edu/abs/2020ARA&A..58..157T} {58, 157}

\bibitem[\protect\citeauthoryear{{Tumlinson} et~al.,}{{Tumlinson}
  et~al.}{2013}]{2013ApJ...777...59T}
{Tumlinson} J.,  et~al., 2013, \mn@doi [\apj] {10.1088/0004-637X/777/1/59},
  \href {https://ui.adsabs.harvard.edu/abs/2013ApJ...777...59T} {777, 59}

\bibitem[\protect\citeauthoryear{{Tumlinson}, {Peeples}  \& {Werk}}{{Tumlinson}
  et~al.}{2017}]{2017ARA&A..55..389T}
{Tumlinson} J.,  {Peeples} M.~S.,   {Werk} J.~K.,  2017, \mn@doi [\araa]
  {10.1146/annurev-astro-091916-055240}, \href
  {https://ui.adsabs.harvard.edu/abs/2017ARA&A..55..389T} {55, 389}

\bibitem[\protect\citeauthoryear{{Turner}, {Schaye}, {Steidel}, {Rudie}  \&
  {Strom}}{{Turner} et~al.}{2014}]{2014MNRAS.445..794T}
{Turner} M.~L.,  {Schaye} J.,  {Steidel} C.~C.,  {Rudie} G.~C.,   {Strom}
  A.~L.,  2014, \mn@doi [\mnras] {10.1093/mnras/stu1801}, \href
  {https://ui.adsabs.harvard.edu/abs/2014MNRAS.445..794T} {445, 794}

\bibitem[\protect\citeauthoryear{{Urry} \& {Padovani}}{{Urry} \&
  {Padovani}}{1995}]{1995PASP..107..803U}
{Urry} C.~M.,  {Padovani} P.,  1995, \mn@doi [\pasp] {10.1086/133630}, \href
  {https://ui.adsabs.harvard.edu/abs/1995PASP..107..803U} {107, 803}

\bibitem[\protect\citeauthoryear{Valentini et~al.,}{Valentini
  et~al.}{2019}]{10.1093/mnras/stz3131}
Valentini M.,  et~al., 2019, \mn@doi [Monthly Notices of the Royal Astronomical
  Society] {10.1093/mnras/stz3131}, 491, 2779

\bibitem[\protect\citeauthoryear{{Villaescusa-Navarro}
  et~al.,}{{Villaescusa-Navarro} et~al.}{2018}]{2018ApJ...866..135V}
{Villaescusa-Navarro} F.,  et~al., 2018, \mn@doi [\apj]
  {10.3847/1538-4357/aadba0}, \href
  {https://ui.adsabs.harvard.edu/abs/2018ApJ...866..135V} {866, 135}

\bibitem[\protect\citeauthoryear{{Wechsler} \& {Tinker}}{{Wechsler} \&
  {Tinker}}{2018}]{2018ARA&A..56..435W}
{Wechsler} R.~H.,  {Tinker} J.~L.,  2018, \mn@doi [\araa]
  {10.1146/annurev-astro-081817-051756}, \href
  {https://ui.adsabs.harvard.edu/abs/2018ARA&A..56..435W} {56, 435}

\bibitem[\protect\citeauthoryear{Wellons et~al.,}{Wellons
  et~al.}{2023}]{10.1093/mnras/stad511}
Wellons S.,  et~al., 2023, \mn@doi [Monthly Notices of the Royal Astronomical
  Society] {10.1093/mnras/stad511}, 520, 5394

\bibitem[\protect\citeauthoryear{{Weltman} et~al.,}{{Weltman}
  et~al.}{2020}]{2020PASA...37....2W}
{Weltman} A.,  et~al., 2020, \mn@doi [\pasa] {10.1017/pasa.2019.42}, \href
  {https://ui.adsabs.harvard.edu/abs/2020PASA...37....2W} {37, e002}

\bibitem[\protect\citeauthoryear{Weng, Peroux, Ramesh, Nelson, Sadler, Zwaan,
  Bollo  \& Casavecchia}{Weng et~al.}{2023}]{weng2023physical}
Weng S.,  Peroux C.,  Ramesh R.,  Nelson D.,  Sadler E.~M.,  Zwaan M.,  Bollo
  V.,   Casavecchia B.,  2023, The physical origins of gas in the
  circumgalactic medium using observationally-motivated TNG50 mocks (\mn@eprint
  {arXiv} {2310.18310})

\bibitem[\protect\citeauthoryear{{White} \& {Frenk}}{{White} \&
  {Frenk}}{1991}]{1991ApJ...379...52W}
{White} S. D.~M.,  {Frenk} C.~S.,  1991, \mn@doi [\apj] {10.1086/170483}, \href
  {https://ui.adsabs.harvard.edu/abs/1991ApJ...379...52W} {379, 52}

\bibitem[\protect\citeauthoryear{{White} \& {Rees}}{{White} \&
  {Rees}}{1978}]{1978MNRAS.183..341W}
{White} S.~D.~M.,  {Rees} M.~J.,  1978, \mn@doi [\mnras]
  {10.1093/mnras/183.3.341}, \href
  {https://ui.adsabs.harvard.edu/abs/1978MNRAS.183..341W} {183, 341}

\bibitem[\protect\citeauthoryear{{White} et~al.,}{{White}
  et~al.}{2012}]{2012MNRAS.424..933W}
{White} M.,  et~al., 2012, \mn@doi [\mnras] {10.1111/j.1365-2966.2012.21251.x},
  \href {https://ui.adsabs.harvard.edu/abs/2012MNRAS.424..933W} {424, 933}

\bibitem[\protect\citeauthoryear{{Zheng} et~al.,}{{Zheng}
  et~al.}{2024}]{2024ApJ...960...55Z}
{Zheng} Y.,  et~al., 2024, \mn@doi [\apj] {10.3847/1538-4357/acfe6b}, \href
  {https://ui.adsabs.harvard.edu/abs/2024ApJ...960...55Z} {960, 55}

\bibitem[\protect\citeauthoryear{{van de Voort}, {Schaye}, {Altay}  \&
  {Theuns}}{{van de Voort} et~al.}{2012}]{2012MNRAS.421.2809V}
{van de Voort} F.,  {Schaye} J.,  {Altay} G.,   {Theuns} T.,  2012, \mn@doi
  [\mnras] {10.1111/j.1365-2966.2012.20487.x}, \href
  {https://ui.adsabs.harvard.edu/abs/2012MNRAS.421.2809V} {421, 2809}

\bibitem[\protect\citeauthoryear{{van de Voort}, {Springel}, {Mandelker}, {van
  den Bosch}  \& {Pakmor}}{{van de Voort} et~al.}{2019}]{2019MNRAS.482L..85V}
{van de Voort} F.,  {Springel} V.,  {Mandelker} N.,  {van den Bosch} F.~C.,
  {Pakmor} R.,  2019, \mn@doi [\mnras] {10.1093/mnrasl/sly190}, \href
  {https://ui.adsabs.harvard.edu/abs/2019MNRAS.482L..85V} {482, L85}

\bibitem[\protect\citeauthoryear{{van den Bosch}, {Ogiya}, {Hahn}  \&
  {Burkert}}{{van den Bosch} et~al.}{2018}]{vdbosch+2018}
{van den Bosch} F.~C.,  {Ogiya} G.,  {Hahn} O.,   {Burkert} A.,  2018, \mn@doi
  [\mnras] {10.1093/mnras/stx2956}, \href
  {https://ui.adsabs.harvard.edu/abs/2018MNRAS.474.3043V} {474, 3043}

\makeatother
\end{thebibliography}


\appendix

\section{Resolution effects on \ion{H}{i} covering fraction of Lyman Limit Systems}

In this section, we discuss the numerical convergence of the \ion{H}{i} covering fraction of LLSs FIREbox. It is well-known today that cold gas in the CGM is not a converged quantity \citep{2023ARA&A..61..131F, ramesh2023zooming}. Recent work shows that increased resolution in the CGM significantly boosts the simulated neutral hydrogen column density and resulting covering fractions \citep[e.g.,][]{2016MNRAS.461L..32F, 2019MNRAS.482L..85V}.

\begin{figure}
    \centering
    \includegraphics[width=\linewidth]{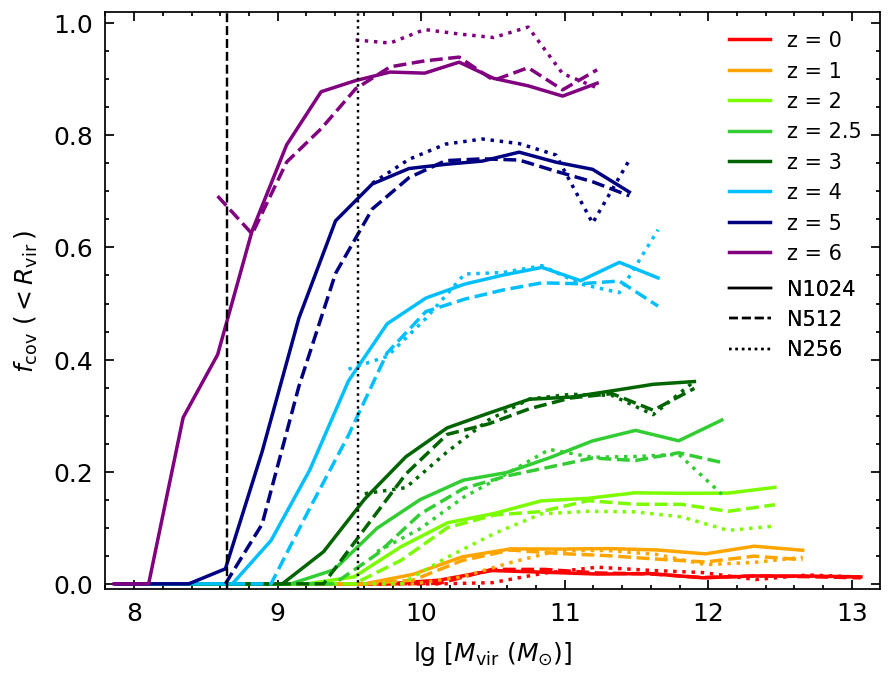}
    \caption{Cumulative covering fraction of LLSs within $\vir$ as a function of halo mass $M_{\mathrm{vir}}$ for different numerical resolutions. Only main haloes were used for this analysis. Different colours indicate different redshifts from $z=0$ to $z=6$ as per the legend. Solid, dashed and dotted lines show results for the N1024 run (fiducial FIREbox run used in this work), the N512 run (8 times lower resolution than N1024) and the N256 run (64 times lower resolution than N1024) respectively. The dashed (dotted) black vertical line indicates the N1024-equivalent lowest halo mass resolved for the N512 (N256) run, corresponding to haloes with at least 168 dark matter particles. We find that the covering fraction slightly increases with resolution, at all halo masses and redshifts. However, this effect is small which suggests that the covering fractions studied throughout this work are close to being converged with resolution.}
    \label{fig:resolution_test}
\end{figure}

We use the lower resolution runs FB512 and FB256 from the \fb{} suite of simulations to investigate resolution effects in our study. The dark matter and baryon masses are 8 and 64 times lower, respectively, and the softening lengths are adjusted accordingly \citep[for details, see][section 2]{2023MNRAS.522.3831F}. We show the cumulative covering fraction within the virial radius $\cov$ as a function of virial mass $M_{\mathrm{vir}}$ for the different runs in Fig. \ref{fig:resolution_test}. We find that the increased resolution of the fiducial N1024 run produces slightly higher \ion{H}{i} covering fractions, at all redshifts and for all masses. This increase is not significant, consistent with findings from \citet{gensior+2023}, wherein it was found that other \ion{H}{i} properties in FIREbox are converged (see Fig. A1 in their online supplementary material). The neutral hydrogen maps used throughout this work (N1024) thus appear effectively converged with numerical resolution.

\section{\ion{H}{i} column density distribution}

\begin{figure*}
    \centering
    \includegraphics[width=\linewidth]{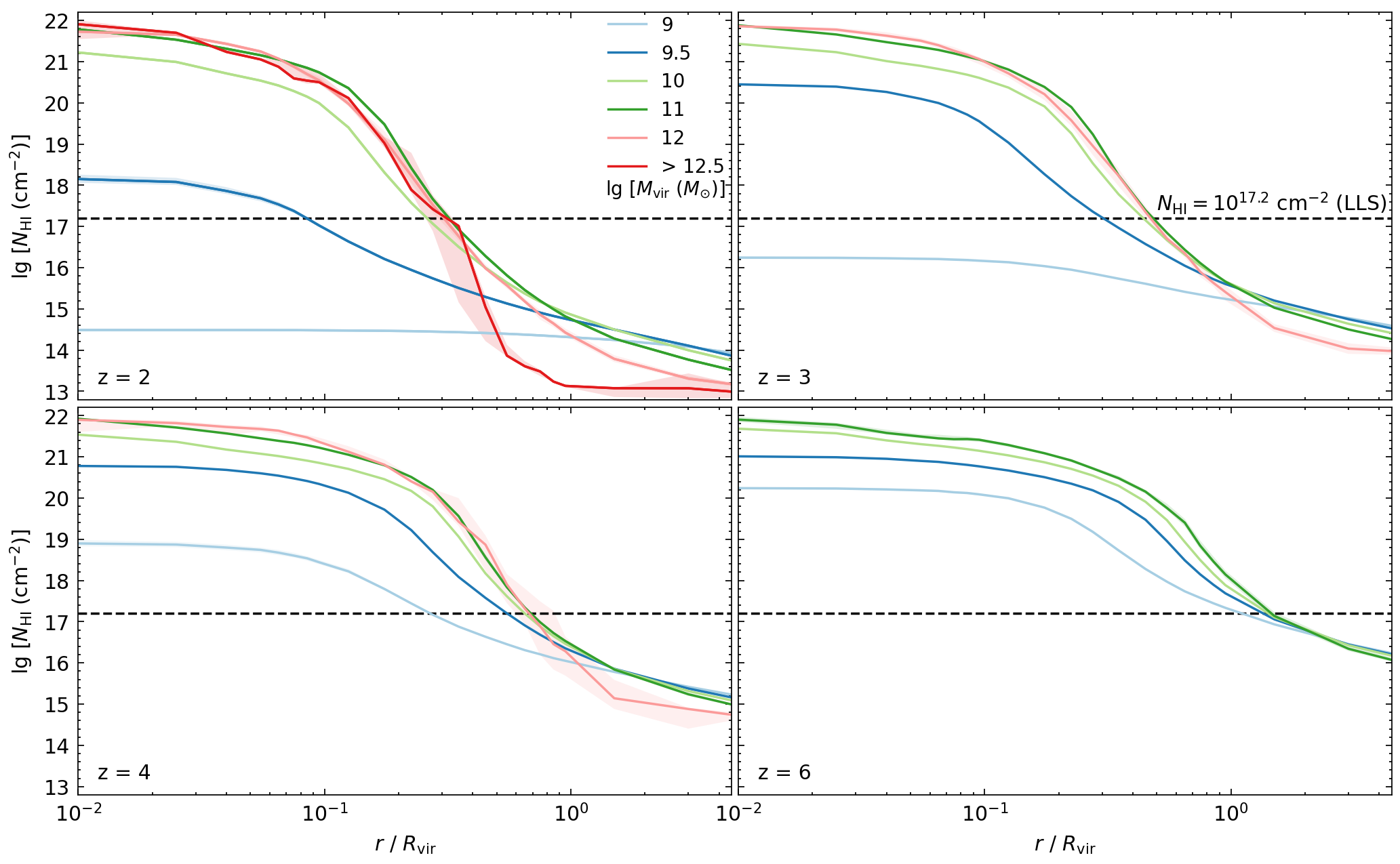}
    \caption{\ion{H}{i} column density profiles of FIREbox haloes as a function of normalised impact parameter for redshifts $z \geq 2$. Solid lines indicate the median $N_{\ion{H}{i}} (r / \, \vir)$ at each impact parameter and the shaded areas show the 5th-95th percentile error on the median obtained from bootstrapping. The mass bins are described in section ~\ref{dfcov}. The redshift depicted in each panel is shown at the bottom left. We indicate the LLS column density cut ($N_{\mathrm{HI}} = 10^{17.2}$ cm$^{-2}$) with a black dashed line. The column density profiles of \ion{H}{i} are similar in the outskirts ($\gtrsim \vir$) for all halo masses. In the inner regions, $N_{\ion{H}{i}}$ increases with increasing halo mass. LLSs are thus less extended in lower mass haloes because the density threshold is reached at lower radii than for more massive haloes.}
    \label{fig:nhi_profile}
\end{figure*}

We show the \ion{H}{i} column density distribution for FIREbox haloes of different masses and for redshifts $z=2, 3, 4$ and 6 in Fig. ~\ref{fig:nhi_profile}. We indicate the neutral hydrogen column density threshold of Lyman Limit Systems ($N_{\mathrm{HI}} = 10^{17.2}$ cm$^{-2}$) by a black dotted line.

We find that the profiles of neutral hydrogen column density are very similar in the outer edges ($\gtrsim \vir$) of \fb{} haloes, for all halo masses and at all redshifts. In the inner regions of haloes, however, we see that the \ion{H}{i} column density strongly increases with increasing halo mass until it reaches a maximum threshold. For massive haloes with $M_{\mathrm{vir}} \geq 10^{11} M_{\sun}$, the density threshold of Lyman Limit Systems is reached at the same normalised radius $r / \, \vir$, further hinting at the scale-invariance discussed in sections ~\ref{dfcov} and ~\ref{sec:fitfct}. While in lower mass haloes ($M_{\mathrm{vir}} \lesssim 10^{10} M_{\sun}$), the density threshold is reached at lower  radii or not reached at all. The LLS differential covering fraction profiles of lower mass haloes are therefore less extended than for more massive haloes.

\bsp
\label{lastpage}
\end{document}